FSM MODELING FOR OFF-BLOCKCHAIN COMPUTATION

by

Christian Gang Liu

Submitted in partial fulfilment of the requirements
for the degree of Master of Computer Science

Dalhousie University
Halifax, Nova Scotia
Feb 2021

# TABLE OF CONTENTS

# LIST OF TABLES

# LIST OF FIGURES

# ABSTRACT

Blockchain benefits are due to immutability, replication, and storage-and-execution of smart contracts on the blockchain. However, the benefits come at increased costs due to the blockchain size and execution. We address three fundamental issues that arise in transferring certain parts of a smart contract to be executed off-chain: (i) identifying which parts (patterns) of the smart contract should be considered for processing off-chain, (ii) under which conditions should a smart-contract pattern to be processed off-chain, and (iii) how to facilitate interaction between the computation off and on-chain. We use separation of concerns and FSM modeling to model a smart contract and generate its code. We then (i) use our algorithm to determine which parts (patterns) of the smart contract are to be processed off-chain; (ii) consider conditions under which to move the pattern off-chain; and (iii) provide model for automatically generating the interface between on and off-chain computation.

# LIST OF ABBREVIATIONS USED

FSM              Finite State Machine

HSM              Hierarchical State Machine

EVM              Ethereum Virtual Machine

IPFS             Interplanetary File System

# ACKNOWLEDGEMENTS

First and foremost, I am grateful to my supervisors, Prof. Peter Bodorik for his invaluable advice, continuous support, and patience during my MCS study. His immense knowledge and plentiful experience have encouraged me in all the time of my academic research and daily life. I would also like to thank Dr. Michael McAllister for approving the research scholarship upon recognizing my achievements during the MCS study and research. I also want to thank Dr. Dawn Jutla, Dr. Qiang Ye and Dr. Vlado Keselj for their time to be the committees of my thesis defense. And they provided valuable feedbacks to my research. It is their kind help and support that have made my study and life in the Dalhousie a wonderful time.

# CHAPTER 1    INTRODUCTION

Blockchain may be considered an immutable and replicated database managed autonomously using a peer-to-peer network and a decentralized replicated virtual machine using distributed timestamping servers (Wikipedia 2020). Supplementarily, a smart contract works like a stored procedure with immutable instructions that are also stored on the blockchain and thus tamper-proof. In general, as the name indicates, a smart contract is typically used to reach and enforce an agreement between engaged parties, wherein that agreement can be modeled and governed by a well-known state machine design pattern of a smart contract (Wöhrer & Zdun, 2018). Many industries have rapidly adopted smart contracts since Ethereum was created in 2015 because smart contract over blockchains is secure, distributed, and manageable.  However, blockchain is still struggling with known issues, such as storage size and scalability, not to mention the nowadays growing Big Data requirements for e-commerce type transactions on blockchains.

## 1.1   RELATED WORK

Blockchain research and software development communities have been developing methods, models, and design patterns to mitigate the aforementioned limitations. For instance, Eberhardt & Tai (2017, 2018) introduced models for off-chain computation and storage.  Off-chain storage is the most prevalent technique as it is easy.  Instead of storing large objects or documents on the blockchain itself, they are stored off-chain while their digital signatures are stored on the blockchain itself.  Any time an object stored off-chain is accessed, its signature is recalculated and checked with the object's signature stored on the blockchain.  Blockchain sharding is another approach to split blockchain data into shards and distribute them among different geographic locations to enhance scalability and reliability (Dang, et al. 2019).

Off-chain computation refers to performing some of the smart contract computation off-chain and using various techniques to verify that the computation results are trustworthy to the level as if they were performed on-chain.  Hybrid off-chaining refers to moving both the storage and computation off-chain (Eberhardt & Tai, (2017, 2018)); however, as it is not clear how one can perform off-chain computation without data (which is on-chain), from now on, unless stated otherwise, we shall use the terms "processing off-chain" or "off-chain computation" to refer to the hybrid off-chaining in which both computation and storage are performed off-chain. Approaches taken to off-chain computation thus far, however, neither address the issue of how to determine which computations should be moved to off-chain, nor how to automate the process systemically – topics of this thesis.

Plasma project makes off-chaining methods the basic building block for smart contract development (Poon & Buterin, 2017). It allows developers to use mainstream programming languages for off-chain computation, of which results are verified by the participants and agreed to by using smart contracts executing on a blockchain. In effect, the blockchain thus serves as a certifier that the distributed off-chain computation participants have verified the computation results and attest to such verification by signing the results.

FSM modeling has been used for modeling Blockchain applications, which is not surprising as blockchains have been referred to as state machines. For instance, Asgaonkar (2018) utilized FSM modeling to introduce a dual-deposit escrow mechanism between Buyer and Seller; AlTawy, et al. (2019) introduced a Blockchain-based delivery system called Lelantos and demonstrated it with FSM modelling; Nguyen, et al. (2018) illustrated his digitizing Invoice system on Blockchain with FSM modeling; Bai & Synnes (2018) proposed an FSM-based reward system for the care of elderlies.

Patterns have also been exploited widely in research and development of smart contracts. For instance, Eberhardt & Tai (2017, 2018) illustrated the Challenge-response pattern and Chess end-game pattern for the off-chain computation, and Mavridou & Laszka (2017) developed a smart contract design pattern for security issues, such as transition counts, locking, and cross-cutting. They used patterns in their search for creating secure smart contracts which were modeled using FSMs that were then transformed into smart contracts.

## 1.2 OBJECTIVES

As the cost of blockchain infrastructure is high, it was quickly determined that one of the "low hanging fruits" in solving this problem is reduction of data stored on the blockchain by storing it off-chain and store on the blockchain its hash-code used to ensure data immutability – a simple, effective, and efficient method. However, moving smart contract processing off-chain is not that simple and research on this topic is in its nascent stage. Our research objectives target the three main issues that need to be resolved by a successful solution and that will be discussed shortly. But before we proceed, we provide a few statements on our terminology.

A smart contract is a collection of methods, which are supported by various data structures, that are invoked by off-blockchain systems. To determine which part(s) of a smart contract should be processed off-chain, it must be determined which collection of methods and supporting data are to be processed off-chain. It is clear that selecting a random collection of methods to be processed off-chain is likely to be less suitable for processing off-chain than a collection of methods that collaborate to achieve a common goal and we shall refer to such a collection of methods as a *pattern*. Hence, when we refer to a *pattern to be processed off-chain*, we are referring to the collection of methods and supporting data to be processed off-chain.

The three issues that we address are:

1. The first issue deals with determining which pattern should be processed off-chain – our first objective. Is one pattern more suitable to be processed off-chain than another, and if so, how do we find a pattern to be processed off-chain?

2. For a pattern to be processed off-chain, it needs to have two properties.

    a. One property deals with the issue of overhead costs. When a pattern is processed off-chain, a trade-off is introduced. A pattern is moved off-chain to reduce the cost of processing it as, it is assumed, that the cost of processing a pattern off-chain is much lower than to process it on-chain. However, there is also overhead cost introduced due to

communication/interaction between on and off-chain processing – hence a trade-off.

    b. Processing a pattern off-chain should not diminish the beneficial blockchain properties, such as resilience to failures or security attacks and immutability of blockchain data.

3. The final issue is how to facilitate interface automatically for interaction between on and off-chain processing.

Loosely speaking, the three issues may be referred to briefly as "What", "When", and "How", respectively: (a) What is to be processed off-chain, i.e., which pattern should be considered for processing off-chain; (b) When processing off-chain should be facilitated, i.e., which conditions need to be satisfied by a pattern to be moved off-chain; (c) and the third one is how to facilitate automatically the interaction between on and off-chain processing.

In addition to tackling the three objectives above, an additional objective was added and to exploit results of (Mavridou & Laszka, 2017), which uses FSMs to model requirements of a smart contract and then translating the FSM model automatically into the smart contract code. Our final objective is to determine if these research results can be used in achieving separation of concerns, when designing a smart-contract pattern that should utilize off-chain processing to reduce its cost: The first concern is satisfying user requirements and then the second concern is processing off-chain.

## 1.3 CONTRIBUTIONS

Before we list contributions, we need to highlight that we build upon previous research results of Marvidou & Lazska (2017(a)(b)), in which application requirements for a smart contract are represented by FSM and then translated into a smart-contract method automatically, for two reasons:

    (i)    One is that graph representation of an FSM enables us to view the connectivity of methods that could be used to analysis in identifying which patterns should be considered for processing off-chain.

(ii)     The second is that we exploit their results in dividing our problem into two sub-problems ((i) satisfying user requirements and (ii) processing off-chain) that are based on separation of concerns, and thus reducing the complexity in that the complexity of each sub-problem is simpler than when solving both problems at the same time.

Our contributions are described for each of the objectives.

1. We developed an algorithm to examine the graph representation of an FSM that models the user requirements in order to find patterns to be considered for processing off-chain: The algorithm looks for patterns that have well-defined goals and minimal connectivity with other patterns, where the connectivity represents interaction across patterns. We applied the algorithm on a selection of real smart-contract use-cases and report on results of its application.

2. We investigated variables that affect the second objective that deals with analyzing a given pattern to determine whether it should be processed off-chain. We broke the problem into two sub-problems:

   a.  A pattern should be processed off-chain only if the trade-off, between the cost benefits/savings outweigh the overhead cost of interaction between the on and off-chain processing. We develop an abstract model to analyze the components of overhead costs and benefits gained in order to evaluate the trade-off. We also show applications of that model in the context of Ethereum smart contracts.

   b.  Besides the benefits-overhead cost analysis, we also investigate the problem of ensuring how to ensure that processing off-chain includes mitigation of the loss of trust in the blockchain solution due to off-chain processing. However, we do not offer concrete solutions as they depend on semantics of the application, semantics that may be obvious to developers but not as obvious for automatic analysis by software.

3. We developed a model that identifies the sub-tasks to be tackled to produce

automatically interface between on and off-chain interaction.

4. Using FSM modeling we successfully applied the concept of separation of concerns by the developer first using FSM modeling to describe the requirements and then producing the smart contract. Only then does the developer, or an expert on processing off-chain, use our approach to determine which smart-contract patterns should be considered to be processed off-chain, under which conditions, and how the on and off-chain processing should interact. If the developer needs to be concerned at the same time about application requirements and patterns to be process off-chain, the problem complexity increases greatly. By using separation of concerns, the developer first creates the smart contract to satisfy the user requirements and only then tackles the issue of processing off-chain and thus simplifying the one complex problem into two simpler ones – albeit still difficult problems.

## 1.4 OUTLINE

The remainder of this paper is organized as follows. Chapter 2 reviews the background to lay out the context for the models and techniques that we use. Chapter 3 explains the FSM modelling approaches for a variety of patterns that often arise in business/commerce. We also develop an algorithm to find patterns for an FSM that represents a smart contract. We then explore properties of an FSM graph representation to find graph properties that can be used to find patterns that are suitable for processing off-chain. We found such graph patterns and describe them and in Chapter 4 we present an algorithm that uses those properties to find patterns that are suitable for processing off-chain.

Chapter 5 investigates how interface for processing a pattern off-chain can be facilitated automatically. We outline informally the protocol for the interaction between on and off-chain processing in terms phases of such interaction and of messages with their content for each phase. In Chapter 6, we address under which conditions should the off-chain processing of a pattern off-chain proceed from two perspectives: One is due to the trade of between cost and benefits of processing a pattern off-chain, while the other is in ensuring that the trust in blockchain properties is not diminished when processing of a pattern is off-chain. Finally, Chapter 7 offers a summary and provides conclusions.

# CHAPTER 2     BACKGROUND

We examine the following topics as background used for our approach:

- FSM Modeling for blockchain applications

- Hierarchical State Machines

- Off-chaining models and approaches

- Smart contract design patterns

## 2.1   FSM MODELING OF BLOCKCHAIN APPLICATIONS

For stateful smart contracts, blockchain researchers frequently model smart contracts using FSMs by describing the blockchain application functions as state transitions. Inputs of the function may be viewed as inputs of FSM transitions. Typically, outputs of the function describe the properties of targeted states. In the blockchain-based application, the present active state is registered as a global variable. Overall, the blockchain application specification is naturally embraced by FSM modeling.

FSM modeling, which is used frequently in many IT domains, includes the following steps:

- Identify the business problem.

- Define the use cases and use scenarios.

- Specify requirements, including actors, states, transitions, inputs, and outputs definitions.

- Identify alternative (exceptional) use cases, such as state timeouts for transitions timeout and system errors.

- Draft FSM for the proposed applications.

- Implement FSM models with smart contracts.

For instance, Mavridou & Laszka (2017) created an FSM model for Blind auction as an example for the proposed FSolidM application (Fig. 2.1).

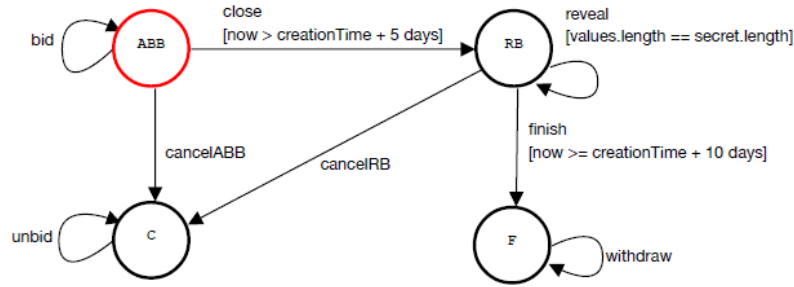

Fig. 2.1 FSM of Blind Auction (Adopted from Mavridou &Laszka (2017))

Mavridou & Laszka (2017) started with use-case analysis by identifying that the bidder, one of the actors, can be in one state from the set of states {*AcceptingBlindedBids, RevealingBids, Finished, canceled*} denoted on the Fig. 2.1 as *ABB, RB, F*, and *C,* respectively. Then inputs, outputs, contract data, and guards are described and *used to convert the FSM description automatically into parameters and return values of smart contract methods.* With the FSM definition combined with the diagram, blockchain developers can easily understand the application design and implement it. We use FSM modeling also to determine which portions of an FSM model can be processed off-chain.

## 2.2    OFF-CHAINING MODELS AND APPROACHES

Eberhardt & Tai (2018) categorized generic models for off-chain computation into three types based on whether it is for storage, computation, or a combination of the two that are performed off-chain.

Off-chain storage: In this model, selected data and transition outputs are moved to off-chain storage, while computation is still on the blockchain. As described before, data is stored off-chain while data's hashcode is stored on-chain, and verification occurs each time the data is retrieved by calculating the hashcode of the retrieved data and comparing it to the hashcode stored on the chain.

Off-chain computation: In this model, FSM pattern states and their transitions are moved to off-chain, while the outputs generated by off-chain computation will be stored and persisted on blockchain once off-chain computation completes. The verification process occurs when the computation moves back to on-chain and storing the results of off-chain computation on the chain – the data needs to be verified by actors affected by the computation and approved/attested by actors as correct results of off-chain computation.

Hybrid Off-chain storage and computation: This model combines the two approaches. That is, computation and data are processed/stored off-chain while results of off-chain computation must be verified and attested by actors. For instance, both computation and storage can be delegated to the sidechain channel Lightning network, with a prudent verification process at the end of the off-chain computation.

We assume that documents and objects are stored off-chain automatically using the hashcodes for verifications, and we concentrate on the Hybrid computation.

## 2.3  HIERARCHICAL FSM

Although FSMs are used frequently in modeling distributed applications and smart contracts, for complex applications, FSM modeling suffers issues of space, scalability, and reusability due to the FSM having only one level of states leading to the flat FSM structure. If there are many states, graphical representation, which is used for ease of representation and understanding, suffers due to the graph's large size. Furthermore, specific patterns are repeated in an application and the flat FSM structure of an FSM graph does not support the re-use of patterns. Therefore, inheritance and encapsulation principles were incorporated into FSM modeling, leading to Hierarchical State Machines (HSMs), first introduced by Harel (1987) and further described and elaborated by Yannakakis (2001). An HSM consists of one or more compound states containing their own internal FSM. Depending on application complexity, designers can now create an internal FSM recursively within an internal state of another internal FSM while leading to Hierarchical FSM layers. We leverage HSM in our modeling of applications and finding which parts of the smart contract represented by an HSM should be performed off-chain.

## 2.4  SMART CONTRACT DESIGN PATTERNS

A smart contract is an executable program used for FSM model implementation. The smart contract can hold complex values, such as struct, defined as FSM properties, and immutable instructions representing state transitions. Design patterns have already been explored for use in the design of smart contracts and a number of examples that we adopt for our work include ((Mavridou & Laszka, 2018), (HeartBank, 2018), (Roan, 2020)):

State Machine: Break down the use case scenario into states to make the use cases transparent and controllable. Our FSM modeling approach naturally adopts this design pattern.

Cross-cutting: Define the timeout modifier and assign this modifier to each state of the transaction because timeout could happen throughout the transaction.

Fail early and Fail loud (HeartBank, 2018): Define failure condition as early as possible in a smart contract logic to prevent continuous expensive on-chain execution.

Upgradable Registry (HeartBank, 2018): Try to separate business logic from the data as much as possible. It encourages Blockchain designers to consider off-chain/sidechain storage. Our approach will potentially maximize its effectiveness.

Locking (Mavridou & Laszka, 2018): This pattern concept is similar to the synchronization mechanism of other mainstream program languages like Java. It locks running smart contract instructions to prevent re-entrancy vulnerability caused by malicious DOS/DDOS attacks.

Transition counter (Mavridou & Laszka, 2018): It enforces transition calls in a planned order. It helps with overcoming transitions reordering issues reported often for blockchains.

Access control (Roan, 2020): This pattern is implemented as the guard in smart contracts' constructor to prevent unauthorized access attempts.

Many smart contract design patterns summarized above need ample patience and prudence from smart contract designers or developers. For instance, smart contract developers need to manually attach the defined modifiers to each required function in terms of the cross-cutting pattern, which might inevitably introduce human mistakes. Fortunately, the FSolidM tool introduced by Mavridou & Laszka (2017, 2018) can automatically implement those patterns as plugins. That is, FSolidM can minimize smart contract developers' efforts for the FSM model transformation into a smart contract and its implementation. We adapt their approach to transforming an FSM model into a smart contract code by incorporating an API to the smart contract to support the smart contract and off-chain processing interaction.

# CHAPTER 3      FINDING PATTERN PROPERTIES

After introducing the notation used to define FSMs for smart contract modeling (section 3.1), we explore patterns that commonly appear and are suitable for off-chaining (section 3.2) and describe how FSMs are used to model them. We studied the properties of the patterns in order to find those that make them suitable for off-chaining. We found and describe such properties and then present an algorithm that uses these properties to identify such patterns in Chapter 4.

## 3.1   FSM Modeling Of Smart Contracts

Blockchains and their smart contracts have states ((Mavridou & Laszka, 2018) and methods that may change the contract state. Consequently, smart contracts lend themselves well to be modeled by FSMs. We model a smart contract with an FSM in the usual way with a set of states, including an initial state, and transitions with input and output. Input transition includes parameters passed to the method, while output, upon exiting transition, includes the new state and other information produced by its computation method.

A smart contract method has computation that can involve state variables stored on the blockchain and therefore implies reading from or writing to the blockchain by reading or writing them. Also, the computation may involve transient values existing only for the duration of the method. Besides, state variables may be simple, such as variables with simple numeric values or enumerated values. Still, they can also be complex objects, such as corporate documents. It is already a standard approach to use off-chain storage techniques for large documents or objects instead of storing them on the blockchain. For instance, one of such methods is Content Addressable Storage Pattern (Eberhardt and Tai, 2017), in which only the hash-code of the object is stored on the blockchain, while the object is stored off-chain, e.g., on a subset of nodes the network (but not on the network itself) or on a sidechain, or in a data system that utilizes hash-code to retrieve data efficiently. The system recalculates the hash code and compares it to the hash code stored on the object retrieved from the database to ensure that the immutability property has not been violated.

As mentioned above, we target smart contracts for business transactions that implicitly represent the workflow between a group of business parties referred to as actors. As a result, the smart contract methods are invoked in response to events that occur by actors' actions. Thus, an FSM F can be described as F = (S, $s_0$, T, I, O), where

- S: Set of n states *{ $s_0$, $s_1$, ..., $s_{n-1}$ }*, where $s_0$ is the initial state of the FSM

- T: Set of transitions *{ $t_i$, i=1, 2, ..., m }*, such that each $t_i$ has two components, $t_{i.from}$ and $t_{i.to}$, that identify the transition to be from the state $t_{i.from}$ to the state $t_{i.to}$, respectively

- I: Set of inputs *{ $x_i$ }*. Each transition $t_i$ has input $x_i$ that causes the transition. ($x_i$ may be a large object or a set of objects)

- O: Set of outputs *{ $y_j$ }*. Each transition $t_i$ has a set of outputs where each $y_i$ may be a single value or it may be an object or a set of objects, whose value is produced by the state activities and is output by the state transition.

We are adapting the method from a previous work (Eberhardt & Tai, 2017), (Mavridou & Laszka, 2018) by developing an FSM model of the smart contract centered on the user requirements and then evaluating the model to decide which subsets of FSM can be implemented off-chain. Our proposed method thus uses the separation of concerns.

To find out which parts of a smart contract should be processed off-chain, we explored many patterns to determine if a pattern is suitable for off-chaining and if so, try to find a property that would assist a developer in finding them. We tried to develop methods for their discovery and, eventually, found properties for patterns that make them suitable for processing off-chain. We describe some of the initial patterns that we considered and algorithms that we developed to identify them; however, our initial algorithms were not general in case of even simple variations in the requirements until we discovered properties that can be exploited by an algorithm to identify them. We describe several such patterns, together with our initial algorithms, in Section 3.2 and then describe the pattern properties that make them suitable for off-chain processing in Section 3.3. We also illustrate these patterns' characteristics, which make them suitable for off-chaining, and, in the next section, we define a general algorithm to find patterns for off-chaining. Our introduction is informal, while in the following chapters, we include a more formal presentation of the whole process. It should be noted that algorithms' efficiency was not considered for the design phase as it is executed infrequently, that is once per smart contract design phase.

In short, while this sub-section (3.2) describes patterns in our initial but unsuccessful efforts in forming a general algorithm to find patterns to be processed off-chain, these efforts led us to recognize specific properties for patterns to be considered for off-chaining computation as described in section 3.3.

## 3.2 Specific Patterns Considered for Processing Off-chain

### 3.2.1 Sequence of Events

Sometimes, in business transactions, activities have to occur in a specific sequence. For instance, Riham (2019) describes a use-case scenario that illustrates a sequence of events occurring in a blockchain-based delivery system model with an FSM shown in Fig. 3.1 as it may occur as a part of a larger use-case (AlTawy, et al. 2017). His proposed delivery system starts with the state in which a buyer places orders for some of the products listed by the merchant. The order document is accepted and digitally signed by merchant. The merchant drops off ordered items at the shipping company. After receiving the package, the shipping company ships it to its destination pickup location. Finally, the buyer picks up the ordered items, and then both the customer and the buyer are provided with the completion of order documentation. An FSM model of this process has shown in Fig. 3.1(a).

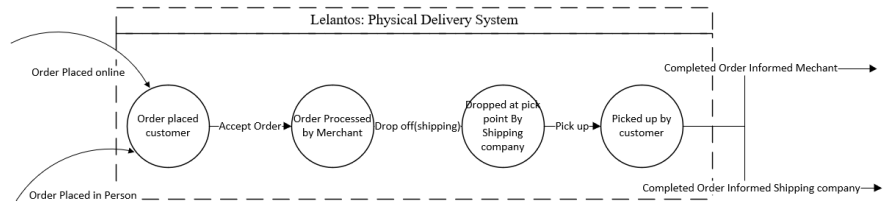

(a) Part of FSM graph showing the Sequence of Events pattern

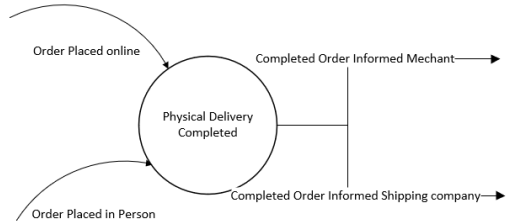

(b) FSM graphs with the subgraph replaced by a node representing HSM

Fig. 3.1    FSM representing a sequence of events pattern.

Such a pattern is suitable for off-chaining as it merely describes activities that need to be executed in a sequence, one after another in a specific order.  The output of each transition in the sequence is the information produced by the state that confirms the completion of the task. Once the exit from the sequence is made, we know the states that have been transitioned, and all outputs from all states in that sequence are collected. On output, during the exit transition out of that sequence, we have all the necessary information about those transitions and can record them on the blockchain.

When it is determined that a pattern, which in this case is a sequence of events represented by an FSM subgraph, should be moved off-chain, the FSM graph is modified, and it is replaced by one node that represents/models a hierarchical FSM (see Fig. 3.1(b) with input being the transition into the initial step in the sequence and output being the exit transition out of the sequence.  Further details will be provided in the next chapter.

Fig. 3.2 illustrates Algorithm 1, which has as input an FSM F, a description of the full FSM using the notation described in section 3.1, and a set S' of nodes that is a subset of the given FSM nodes graph. Algorithm 1 determines whether or not the subset S' of S forms a Sequence of Events pattern.

---
**Algorithm 1:** Algorithm to retrieval subgraph of Specific Sequence of Events
---

**Input: Tested** $S$, **where** $S$ **is from FSM F = (**$S$, $T$, $I$, $O$**)**

**Output:** Return subgraphs for matching Sequence of Events pattern, otherwise return false

(1)   $matched = \{\}$

    // T is the set of transitions of FSM F with states S

(2)   **Function** getSequenceOfEventSubgraph( $T$ ):

       **Set** $remainderOfT = T$

(3)     **if** $T'$ is not empty **then:**

(4)         **for** $t$ **in** **remainderOf** **T do:**

(5)            $Scurrent = t.to$

(6)            $S_{previous} = t.from$

(7)            **if** ($S_{current} \in S$ **and** $S_{previous} \in S$) **or** ($S_{current} ==$ **NULL and** $S_{previous} \in S$) **or** ($S_{current} \in S$ **and** $S_{previous} ==$ **NULL) then**

(8)               **if** $S_{previous} ==$ **NULL and** $S_{current} ! =$ **NULL and** $matched$ **is empty then**

(9)                 $matched = matched$ ∪ $S_{current}$

(10)                $remainderOf T = remainderOfT - t$

(11)                getSequenceOfEventSubgraph ($remainderOfT$)

(12)               **end**

(13)               **if** $matched$ **is not empty and** $S_{previous} == matched.last$ **then**

(14)                **if** $S_{previous} ! =$ **NULL and** $S_{current} ! =$ **NULL then**

(15)                  $matched = matched$ ∪ $S_{current}$

(16)                 $remainderOf T = remainderOf T - t$

(17)                 getSequenceOfEventSubgraph ($remainderOfT$)

(18)                **end**

(19)               **if** $S_{previous} ! =$ **NULL and** $S_{current} ==$ **NULL then**

(20)                 $remainderOf T = remainderOfT - t$

(21)                 getSequenceOfEventSubgraph ($remainderOfT$)

(22)     **else**

(23)         **if** $matched == S$ **then**

(24)            **Return** $matched$

(25)     **else**

(26)         **Return** $false$

---

Fig. 3.2 Algorithm 1 to identify the sequence of events pattern for given subgraph S'

The algorithm above utilizes the following properties to identify the pattern:

- The subgraph must form a chain (linked list) of n nodes.

- The first node, referred to as an entry node, may have one or more incoming edges and precisely one outgoing edge.

- The last node in the chain must have one incoming edge and one or more outgoing edges.

We should add that the algorithm is unusable in real situations as it finds only "chain" patterns and not minor variations to it. For instance, if we modify the pattern slightly by introducing a couple of states denoting shipment either by air or ground to the pickup point (the amended pattern is shown in Fig. 3.3), Algorithm 1 above will fail to discover it even though it is only a minor variation of the pattern in Fig. 3.1(a). We need a more flexible algorithm that would also find variations in the sequence of events.

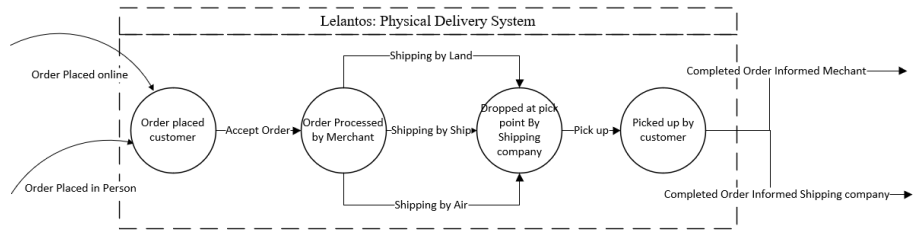

Fig. 3.3 Pattern that is a minor variation of the pattern in Fig. 3.1 (a)

### 3.2.2  Two-parties Agreement Pattern

There is a frequent need for two parties/actors to agree on something in contractual dealings, such as an agreement on a contract. Consider an FSM representing such activities in a smart contract. Aditya (2018) provides an example of a Dual deposit escrow pattern on the blockchain represented by an FSM in Fig. 3.4. Both the buyer and the seller make a security deposit upon reaching a purchase agreement. Whether it is the buyer or the seller, who makes the first deposit, does not matter. As it may arise in an FSM modeling of some smart contracts, such a pattern is shown graphically in Fig. 3.4(a).

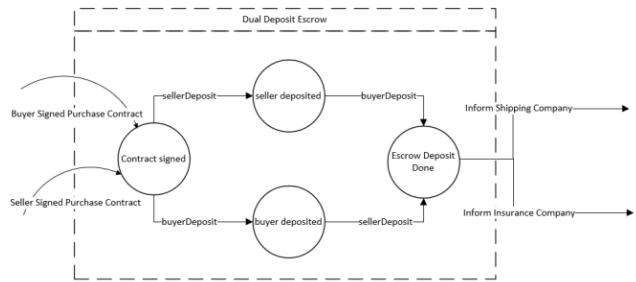

(a)  A pattern, in an FSM graph, representing a two-party agreement

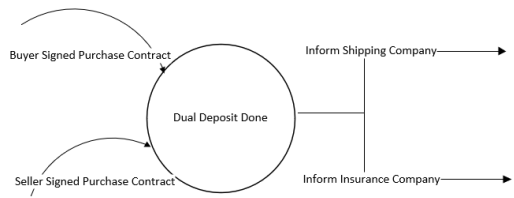

Replacement of 3.4 (a):  a two-party pattern with a node representing an HSM

Fig. 3.4    FSM graph representing a two-party agreement

*Two Parties' Agreement pattern* is also suitable for off-chaining as it describes two parties who wish to agree on something or share the same purpose. Both parties receive the same input. And the output, which usually is a document with both parties' digital signatures, will serve as an input for the next state. When it is determined that a pattern, which is a two-party agreement, should be moved off-chain, the FSM graph is modified by replacing the subgraph with a node representing an HSM (see Fig. 3.4(b)).

For the rest of the patterns, we found it convenient to write a function to find the start and end nodes of a subgraph and to also determine that it has certain properties, namely *DetermineStartAndEndNode*(T', S'). The function's input is subgraph S'. Output identifies start and end nodes, if found, otherwise it returns False Boolean value if not found. The function is used by the algorithms to find remaining patterns we discuss, patterns *Two Parties' Agreement, Any One of N,* and *M out of N* pattern.

```
/*  identify the start node and end node*/
Input: (Ts, Ss ) from the FSM F' = (Ss, Ts, Is, Os)
   Output: Identified initialNode and endNode, otherwise Boolean False if start node and end node can not
   be found.
1 Function DetermineStartAndEndNode(Ts,
        Ss):
2      foundStartNodeEdges = 0
3      foundEndNodeEdges = 0
4      isSameEndNode = true
5      isStartEndNode = true
6      for t in Ts do
7          Scurrent = t.to
8          Sprevious = t.from
9          if Sprevious! = NULL and Scurrent == NULL then
10             if (endNode! = NULL) and endNode! = Sprevious ) then
11                 isSameEndNode = false
12             end
13             else
14                 endNode = Sprevious
15                 foundEndNodeEdges = foundEndNodeEdges+1
16         if Sprevious == NULL and Scurrent! = NULL then
17             if (initialNode! = NULL) and initialNode! = Scurrent ) then
18                 isSameStartNode = false
19             end
20             else
21                 initialNode = Scurrent
22                 foundStartNodeEdges = foundStartNodeEdges+1
23         if (foundEndNodeEdges == N) and (foundStartNodeEdges == N) and
            (isSameStartNode == true) and (isSameEndNode == true) then
24             Return ( initialNode, endNode )
25         end
26         else
27             Return False
28     end
29 return
```

Fig. 3.5  Reusable Function to Determine the Start and End Nodes

Again, the function in Fig. 3.5 finds the start and end nodes/s states of a pattern. A start node of a pattern is a state from which we start to test a subgraph S'. End of a pattern is such that it has an end node representing end of testing a subgraph S'. We use this function to find the start and ends when we decide if a subgraph is one of *Two Parties*' patterns, *Any of N* and *M out of N* pattern. It identifies possible N branches between the start node and end node (line 9-22). Moreover, the number of branches that successfully reach the end node has to be more than the required approvals (line 23-24). In other words, we are looking for a pattern that starts with branching and an end node is identified by the found branches merging together at the same "end" node. If it is a 2-way agreement, we are looking for two branches that merge, and for an n-way agreement, we search for n branches meeting in one node, the end node.

Algorithm 2, shown in Fig. 3.6, finds a two-party agreement pattern. However, the problem is that it only finds the specific pattern shown in Fig. 3.5, but it does not find any similar patterns that may also be suitable for off-chaining. For instance, Fig. 3.7(a) shows an FSM model of a similar mortgage approval pattern. The request may be coming from a real estate agent on behalf of the buyer or the buyer herself. To approve the mortgage, the property land claim has to be surveyed and also a property appraisal has to be performed that is then followed by obtaining an insurance quote (to confirm insurability). This pattern is similar to that of the two-party agreement. It is also suitable for off-chaining as long as the off-chain computation is completed, the approved mortgage and relevant documents (reports from the appraisal, insurance quote, buyer's financials, and the approved mortgage) are saved on the blockchain.

**Algorithm 2:** Algorithm to get subgraph of Two Actors Pattern

    **Input:** $(T_s, S_s)$ **from the FSM F' = $(S, T, I, O)$**
    **Output:** subgraph *matched* of Two Actors pattern, otherwise false boolean vlaue if it is not matching

(1) **Import DetermineStartAndEndNode from Algorithm 2**
(2) *branches* ? An **Array** of **stack**, *branches.length* = 2
      /* defining recursive function to scan both T and S                                        */
(3) **Function getTwoPartiesSubgraph($T'$, $S'$):**
(4)     SET *remainderOfT* = $T'$
(5)     SET *remainderOfS* = $S'$
      /* add initialNode to each branch of *branches* as the first element. return false if initialNode is NULL    */
(6)     **if** *(initialNode != NULL)* **and** *(endNode != NULL)***then**
(7)       **for** *oneBranch* **in** *branches* **do**
(8)         *oneBranch.push(initialNode)*
(9)     **else**
      /* else call algorithm 2 to initlize start and end nodes                            */
(10)       ( *initialNode, endNode* ) = DetermineStartAndEndNode $(T_s, S_s)$
(11)     **end**
(12)     **for** *t* **in** *remainderOf T* **do**
(13)       $S_{current}$ = *t.to*
(14)       $S_{previous}$ = *t.from*
(15)       **for** *s* **in** *remainderOf S* **do**
(16)         **if** $s == S_{current}$ **then**
(17)           *remainderOfT* = *remainderOfT* - *t*
(18)           *remainderOfS* = *remainderOfS* - *s*
(19)           **for** *oneBranch* **in** *branches* **do**
(20)             **if** *oneBranch.peek()* == $S_{previous}$ **then**
(21)               *oneBranch.push(s)*
(22)               getTwoPartiesSubgraph (remainderOfT, remainderOfS)
      /* get matched subgraph, or false value.                                      */
(22)   **// if the size of** *branches* **equal to 2, it means the pattern is two parties pattern**
(23)   **If Length of** *branches* **== 2**
(24)     **Set** *matched* = **UNION of** *branches*
(25)     **Return** *matched*
(26)   **Else**
(27)     **Return False**
(28) **return**

Fig. 3.6 Algorithm 2 to identify the Two-Parties pattern for given subgraph F'

Algorithm 2 in Fig. 3.6 is used to identify the subgraph of the Two-Parties pattern. It first identifies the start node and the end node (line 10). Then the algorithm pushes the start node as the first element to both the branches representing two parties, respectively in the FSM subgraph. The function *getTwoPartiesSubgraph* recursively pushes the current scanned nodes connecting with the previous nodes (line 12-22). These previous nodes must be the same as the last identified elements for each of the branches. At the end of Ts and Ss scanning, two sets of states presenting two parties in FSM should have the last elements equal to the same end node and return matched subgraph. Otherwise, the algorithm returns false to indicate that the examined subgraph does not match the *Two-Parties agreement pattern* (line 23-27).

Another small variation of the graph of Fig. 3.7(a) is shown in Fig. 3.7(b) – additionally, the buyer's financials must be reviewed to ensure that the buyer has enough current and potential future funds to carry the mortgage. Again, our Algorithm 2 does not find it due to the slight variations in the FSM graph. Clearly, we need an algorithm that not only finds the *Two-party agreement pattern* but also its variations, such as those in Fig. 3.7.

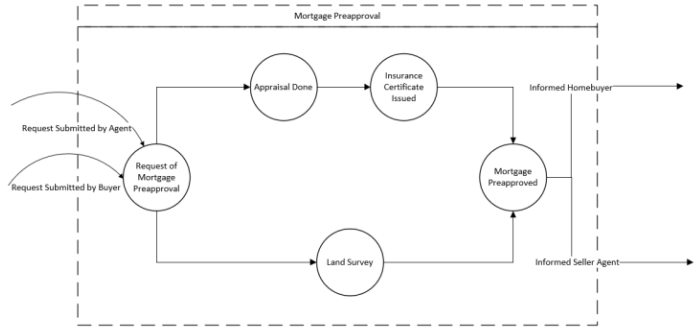

(a) Pattern that is similar to that of Fig. 3.4(a)

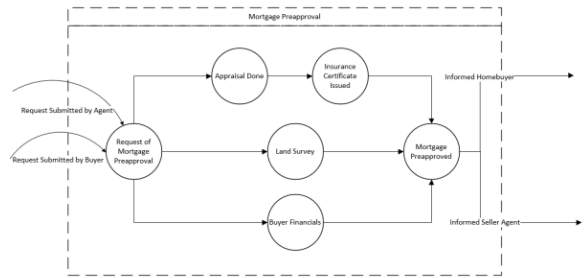

(b) Pattern that is a slight variation of that in Fig. 3.7(a)

Fig. 3.7 Patterns for mortgage approval

### 3.2.3 Permission by Any One of N

In some situations, an agreement is needed by *Any One of N* actors. For instance, an inspection of some items may be performed by any one of three inspection companies (e.g., based on availability for inspection). An FSM is shown in Fig. 3.8(a), representing such activities in a smart contract representing one of three actors' permission. For instance, regarding the purchase agreement and contract, the FSM contains four states { *Offer Accepted, Inspector A inspecting, Inspector C refused, Inspector B Inspecting* }. The corresponding Inputs are { *acceptedOffer, buyerSignedContract, sellerSignedContract, FinalizedContract* }. Note that input of the subgraph of FSM is a house offer document, followed by three states, each representing requests for inspection sent to three inspectors { *Request to Inspector A, Request to Inspector B, Request to Inspector C*}. However, only one inspection submission is required to proceed with the rest of the transitions. As a consequence, only one of the inspectors needs to submit the final inspection report. This FSM subgraph's output is the Final Inspection Report submitted by any one of the three inspectors.

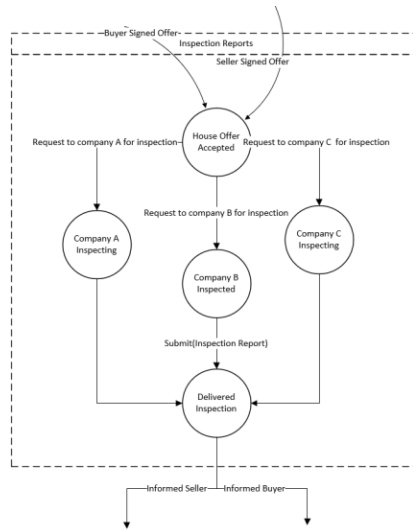

(a) Part of FSM graph representing work performed by anyone of a set of three actors (inspection companies)

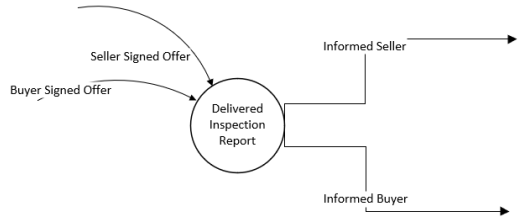

b)  FSM of the part (a) above with the subgraph replaced by a node representing HSM

Fig.  3.8   FSM graph representing an Any One of N Pattern

Again, intuitively, such a pattern is suitable for off-chaining as long as we capture on the blockchain the proof that one of the actors provided inspection.  If the developer decides that the sub-graph showed in Fig. 3.5(a) should be moved off-chain for execution, we represent it by replacing the subgraph with a Hierarchical FSM, shown as one node Fig. 3.8(b).  The node that replaces the subgraph represents a Hierarchical FSM.

We describe Algorithm 3 below used to identify *Any One of N* pattern by examining the FSM subgraph. It firstly identifies the start node and the end node (line 12). The algorithm pushes the start node as the first element to a FILO stack representing a diagram's visual branch. *getMoutNSubgraph* (line 5-29) recursively pushes the current scanned nodes connecting with the previous nodes. That is, these previous nodes must be the same as the last identified elements of branches. At the end of Ts and Ss scanning, all the branches should have the last elements equal to the same end node. Otherwise, the algorithm returns false to indicate that the examined subgraph does not match *Any One of*

*N* pattern (line 25-29).

| Algorithm 3: Obtain Subgraph of Any—M—N of N Pattern |
|---|

**Input:** FSM **F** = ($S_s$, $T_s$, $I_s$, $O_s$)
  Tested Subgraph F' = (**S'**, **T'**, **I'**, **O'**) is the subset of FSM F
  Least of approval $M$ = 1
**Output:** Subgraph *matched* of Any—M—N of N Pattern or false boolean value if it is not matching

(1) **Import DetermineStartAndEndNode from Algorithm 3**
(2) *branches* ← An **Array** of **stack**, *branches.length* = M
(3) *remainderOfS* = S'
(4) *remainderOfT* = T'
    /* defining recursive function to scan both T and S                                              */
(5) **Function** getMOutNSubgraph(*remainderOfT*,*remainderOfS*):
    /* add initialNode to each branch of *branches* as the first element. return false if initialNode is NULL */
(6)   **if** *(initialNode != NULL)* **and** *(endNode != NULL)* **then**
(7)     **for** *oneBranch* **in** *branches* **do**
(8)       *oneBranch.push*(*initialNode*)
(11)  **else**
      /* else call algorithm 3 to initlize start and end nodes                                       */
(12)    ( *initialNode*, *endNode* ) = DetermineStartAndEndNode ($T_s$, $S_s$)
(13)  **for** *t* **in** *remainderOf* T **do**
(14)    $S_{current}$ = *t.to*
(15)    $S_{previous}$ = *t.from*
(16)    **for** *s* **in** *remainderOf* S **do**
(17)      **if** *s* == $S_{current}$ **then**
(18)        *remainderOfT* = *remainderOfT* − *t*
(19)        *remainderOfS* = *remainderOfS* − *s*
(20)        **for** *oneBranch* **in** *branches* **do**
(21)          **if** *oneBranch.peek*() == $S_{previous}$ **then**
(22)            *oneBranch.push*(*s*)
(23)            getMOutNSubgraph (remainderOfT, remainderOfS)
    /* get matched subgraph, or false value.                                                         */
(24)  **// if the size of** *branches* **equal to N, it means the pattern is any or M out N pattern**
(25)  **If** Length of *branches* == *N*
(26)    **Set** *matched* = **UNION of** *branches*
(27)    **Return** *matched*
(28)  **Else**
(29)    **Return False**
(30) **return**

Fig. 3.9 Algorithm 3 to identify an *M out of N* Pattern

Algorithm 3, depicted in Fig. 3.9, supports the following:

- A variety of states for each actor: In order words, the pattern allows different numbers of states and transitions in the branch's middle and

- Various amounts of approvals: That is, approval, as a field variable, is defined as the number of approvals needed to reach the end of the pattern. This algorithm also works with M out of N and N out of N patterns with configurable values of approvals in the algorithm.

For the approval of *Any One of N*, we use a parameter value of 1; for approval of *M out N*, we use a parameter value of M; and for the approval of *N out of N*, we use a parameter value of N.

### 3.2.4 Agreement of M out of N (M/N)

In some situations, an agreement is needed by a minimum M of N actors before any further activities occur. For instance, an agreement by a majority may be required before proceeding. Fig. 3.10(a) shows a sample FSM diagram representing a situation in which an agreement is required by two out of three actors for the Hyperledger consensus mechanism. An update needs to be approved by at least 2/3 +1 nodes of the whole blockchain network during its propagation process.

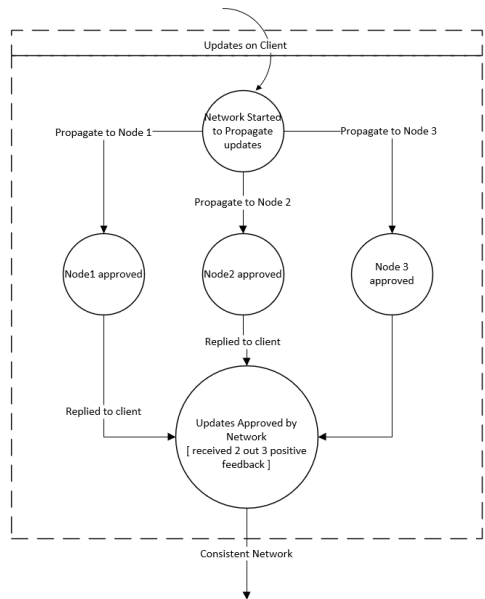

(a) Part of FSM graph representing permission needed from at least two of a set of three actors

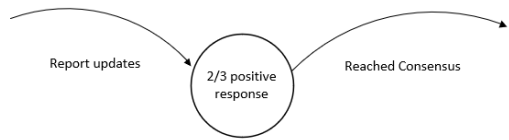

(b) Graphs as in Fig. 3.10(a) with the subgraph replaced by a node representing HSM

Fig. 3.10   FSM graph representing an *M out of N* Pattern

It should be noted that the previously discussed patterns *Any One of N* and *Two Parties' Agreement*, discussed in previous chapters, are just special cases of the "*Agreement of M out of N (M/N) pattern*". As discussed in *Any One of N pattern*, Algorithm 3 works with *M out of N* if we set the approval value as M.

### 3.3   PROPERTIES OF SUBGRAPHS FOR OFF-CHAINING

Recapping, we explored and examined patterns that are suitable for off-chaining with the objective to find a method to recognize such patterns automatically. We also developed an algorithm for each pattern, such as Algorithm 1, 2 and 3 defined above. We also observed that these algorithms were not applicable for actual use. They were specialized to find specific patterns based on properties specific to each pattern: Thus, we reviewed the patterns while searching for some properties common to these patterns. And, examining subgraphs shown in figures 3.1 to 3.9, we found such properties:

- There is only one node/vertex (representing a state) with any incoming edges from any external nodes to any internal nodes of the subgraph – such a node is referred to as an entry node and represents an entry state of the subgraph.

- There is only one vertex/node (representing a state) with any outgoing edges from an internal node of the subgraph to any external node – such a node is referred to as an exit node and represents an exit state.

- With the exception of the entry and exit nodes, all subgraph vertices have edges that connect only to the subgraph's nodes referred to as internal nodes.

Subgraphs of an FSM graph that satisfy the above properties are referred to as simple subgraphs. That a subgraph has only one entry and one exit node mean that once the subgraph's initial state is reached, it has only one specific purpose(s) reached upon transition from an exit vertex of the subgraph. Consequently, the subgraph FSM behavior is determined by the input provided by the transition that is an entry to the subgraph and subsequent events with their inputs. If the subgraph is processed off-chain and all events, inputs, and outputs of transitions are recorded in output upon exit from the subgraph (executed off-chain) – and can all be recorded on the main blockchain itself. Hence, the blockchain will contain all information required for determining the transitions and associated information (inputs/outputs) for the off-chain execution. Of course, depending upon the application's semantics, not all inputs/outputs and transitions need to be recorded on the main blockchain – we shall discuss this in a later chapter.

# CHAPTER 4    ALGORITHM TO FIND PATTERNS

We analyzed several patterns found in research papers. Those patterns include, but not limited to, *Fail early and Fail loud, State machine, Upgradable registry, Transition counter* and others, as well as patterns to alleviate different protection problems, such as the pattern of *locking* to avoid re-entry and access patterns in smart contract programming ((Mavridou & Laszka, 2018), (HeartBank, 2018), (Roan, 2020)).  We studied the models for coordination of activities amongst separate groups, such as the pattern for M/N actors' approval and its sub-variants, two parties/actors' approval, any actor's approval and all actors' approval. We also studied patterns to check for graph properties that could be used by an algorithm to locate patterns in a smart contract that can be processed off-chain. We postulated that a pattern is suitable to be processed off-chain if it has a single objective or a few localized objectives, and its consequences impact identifiable entities.

In the previous chapter, we described properties of subgraphs of an FSM graph representing the smart contract being developed, that can be used by an algorithm to find subgraphs suitable for processing off-chain. In this chapter, we describe such an algorithm and thus are answering the question of "What" (to process off-chain). In the subsequent chapters, we assist the developer in providing the interface between on and off-chain computation (and thus answering "How") and then we tackle the issues of deciding whether or not a simple subgraph should be processed off-chain (answering "*When to off-chain*").

In this chapter, we built an algorithm that analyzes the FSM graph defined by the set of states denoted as S, to find patterns to be considered for processing off-chain. To do so, we model a smart contract with an FSM by utilizing the notation in (Mavridou & Laszka, 2018). Although that notation was already presented in a previous chapter, we repeat it here, together with the definition of a Hierarchical FSM, for the readers' convenience:

An FSM F can be described as F = (S, s0, T, I, O), where

- S: Set of n states *{ $s_0$, $s_1$, …, $s_{n-1}$ }*, where $s_0$ is the initial state of the FSM

- T: Set of transitions *{ tᵢ, i=1, 2, …, m },* such that each $t_i$ has two components, ti.from and ti.to, that identify the transition to be from the state ti.from to the state ti.to, respectively

- I: Set of inputs *{ xᵢ }.* Each transition $t_i$ has input $x_i$ that causes the transition. ($x_i$ may be a large object or a set of objects)

- O: Set of outputs { $y_j$ }. Each transition $t_i$ has a set of outputs where each $y_i$ may be a single value or it may be an object or a set of objects, whose value is produced by the state activities and is output by the state transition.

We adopted the HSM definition that is based on induction from (Yannakakis M, 2000). We assume that S is a connected graph representing an FSM that does not have any unreachable states. An FSM can be treated as a particular form of hierarchical machine. Suppose that M is a set of HSMs. If F is an FSM with a set of states, S, and a mapping function f: S ↦ M, then the triple (F, M, f) is an HSM. Any HSM has a corresponding equivalent FSM that can be achieved by "flattening" the hierarchy in HSM: Each state denoted as s ∈ S that represents an HSM is replaced by its mapping (f(s)). HSMs understand the same expression as their respective flat FSMs — HSMs do not improve FSM expressiveness, only briefness.

Before we describe the algorithm to find patterns to be considered for processing off-chain, we briefly review the context for using it. Once the algorithm is used to find the simple subgraphs, which are ordered and presented to the developer one after another for deciding whether they should be executed off-chain: If a simple subgraph pattern is to be processed off-chain, then the FSM model is modified to represent the off-chain processing. Firstly, we describe an algorithm in the section 4.1 for detecting subgraphs with desirable off-chaining properties to find all the simple subgraphs. Then, we describe its application on actual use case examples in the section 4.2.

## 4.1 Algorithm To Find Simple-Subgraphs Patterns

The algorithm, shown in Fig. 4.4, finds simple subgraphs that possess the previous chapter's graph properties. It is a straightforward brute-force algorithm for simplicity in presentation and ease of understanding rather than concentrating on efficiency and formalism. Furthermore, the algorithm is used in the design phase only, which is a one-time task, rather than executing a smart contract once it is designed and developed. Recall that simple subgraphs contain only one entry and one exit node. An entry node may have incoming edges from nodes outside the subgraph in addition to edges to other nodes of the subgraph. Similarly, an exit node may have outgoing edges to nodes not members of the subgraph.  We call such subgraphs as simple subgraphs.  A simple subgraph is suitable for off-chaining as once a transition transits into the entry node, its processing is "internal" to the subgraph until a transition from the exit node to a node outside the subgraph.  Thus, if a simple subgraph is processed off-chain, once the execution transits to the simple subgraph's entry node, all subsequent calls to the smart contract method are processed off-chain until the state of execution transits to the subgraph's exit node, after which all subsequent smart contract method invocations are processed back on the blockchain.

## 4.2  HELPER FUNCTIONS

Before we proceed with the algorithm, we describe several helper functions for which it is assumed that the given FSM F is global and hence accessible by them.  Consequently, the set of states denoted as S,  of the FSM F, is also global and accessible by the following functions and the algorithm.

set *generateAllSubgraps* (parameter S) … a method that has, as input, FSM F = (S, s0, T, I, O) and returns a set of elements where each element is a subgraph of S. The function *genrateAllSubgraphs* (Fig. 4.1) is a reusable function that finds all possible subsets of S.

---

**Input: S**
**Output:** All Possible Subsets $X$ of S
(1)  **Function** generateAllSubgraps(S):
(2)       Set $X$  = NULL;
(3)       Set $size_{min}$ = 2;
(4)       set $size_{max}$ = S'.size;
(5)       // recursively loop all possible sizes of set to fill in all possible combination.
(6)       for $size$ in range($size_{min}$,  $size_{max}$):
(7)            recursively push any possible combinations of $size$ states into $X$
(8)       return $X$;
(9)  **return**

---

Fig.  4.1 Helper function *genrateAllSubgraphs* to find all possible natural subsets of S

boolean *isSimpleSubgraph* (parameter: set S') … the method determines whether S'⊂ S satisfies the simple subgraph properties. If the result is positive, then the tested subgraph is a simple subgraph. Otherwise, the tested subgraph is not a simple subgraph.

```
Input: S'
Output: Boolean Value of Identification of Simple Subgraph
(1)   Function isSimpleSubgraph(S'):
(2)      // check three properties of simple graph
         // check only start node has all incoming edges
(3)      S_start = NULL
(4)      for each s in S' {
(5)         Set s_from  =  the state all incoming edges come from
(6)         If s_from is not in S' {   // if the from state is outside of the pattern
(7)            Set S_start  = s;
(8)            }
(9)      }
(10)     If S_start == null {
(11)        Return False;
(12)     }
         // check only start node has all outgoing edges
(13)     S_end = NULL
(14)     for each s in S' {
(15)        Set s_to  =  the state all outgoing edges go to;
(16)        If s_to is not in S' {  // if the to state is outside of the pattern
(17)           Set S_end  = s;
(18)           }
(19)     }
(20)     If S_end == null {
(21)        Return False;
(22)     }
         // check all other nodes only connect to internal nodes.
(23)     for each s in S' {
(24)        If s != S_start AND s != S_end {
(25)           Set s_to  =  the state that outgoing edge of s goes to;
(26)           Set s_from  =  the state that incoming edge of s comes from;
(27)           If s_to is not in S' OR s_from is not in S' {
(28)              Return False;
(29)              }
(30)           }
(31)     }
(32)     return true
```
Fig. 4.2  Helper function *isSimpleSubgraph* to identify Simple Subgraph from giving subgraph S'

boolean *isProperSubgraph* ( parameters: set S1, set S2) … method returns true if S1 ⊂ S2. The function *isProperSubgraph* (Fig 4.3) checks if one subgraph is the subset of another subgraph. The proper subgraph identification relies on the positive response of the function.

```
Input: S1, S2
Output: Boolean Value of Identification of S1 is the proper subset of S2
(1)   Function isProperSubgraph:
(2)      for S' in S1:
(3)         if S'  is not in S2 :
(4)            return False;
(5)      return True;
```
Fig. 4.3  Helper function *isProperSubgraph* to identify subgraph S1 is the proper subset of another subgraph S2

## 4.2.1   Algorithm to find simple subgraphs

Equipped with the three helper functions, we defined Algorithm 4 in Fig. 4.4 to find all the simple subgraphs in a graph representation of an FSM F.

| Algorithm 4: find all simple subgraphs for an FSM F |
|---|
| **Input: FSM F = $(S_s, T_s, I_s, O_s)$** |
| **Output:** set $Y = S1, S2, \ldots, Sn \ldots$ set of all simple subgraphs $S1, S2, \ldots, Sn$ of FSM $F$. Each $Si \in Y$ has an attribute $Si.count$ that is the number of other simple subgraphs $Sj, i \neq j$, in $Y$ that are subsets of $Si$, i.e., where $Sj \in Si$ |
| (1) **// Any set $Z$ may have an attribute $Z.count$** |
| (2) **Set $X$ = generateAllSubgraps (FSM $F$ );** |
| (3) **Set $Y = NULL$;** |
| (4) **For each subgraph $S$ in $X$** |
| (5) **If isSimpleSubgraph ($S$) then $\{S.count = 0; Y.add(S)\}$** |
| (6) **// Attribute $S.count$ denotes how many subgraphs $S$ has** |
| (7) **For each $Sx$ in $L$** |
| (8) **For each $Sy$ in $L$** |
| (9) **If isProperSubset ($Sy$, $Sx$) then $Sx.count = Sx.count+1$;** |

Fig. 4.4 Algorithm 4 to find all simple subgraphs for a give FSM F

After all subgraphs are generated (line 2 in Algorithm 4), each one is determined whether it is a simple subgraph or not. If it is, it is stored in a set of simple subgraphs (line 5). Line 6 finds out for each simple subgraph how many other simple subgraphs it contains. It exploits the property that for any two simple subgraphs of one FSM graph, the two simple subgraphs are such that either (i) one is a proper subset of the other or (ii) the two subgraphs S1 and S2 share at most one node and that that node has particular properties that are the result of Theorem 1. We use the properties to count for each simple subgraph how many other simple subgraphs it has as its subgraphs. We use it to order the found simple subgraphs by the *S.count* property and present to the developer simple subgraphs in sorted order by that count (of other simple subgraphs it has) when the developer is deciding for each simple subgraph whether it should be processed off-chain. That is, the developer is asked to decide whether to process S1 off-chain before deciding on S2, if *S1.count* is higher than *S2.count*.

We now present Theorem 1 to describe relationships between simple subgraphs generated for one FSM.

*Theorem 1*: Consider an FSM F, with a set of states, S, that has two simple (connected) subgraphs, S1 $\subset$ S and S2 $\subset$ S, where |S1| >= 1 and |S2| >= 1. Then one of the following holds:

a) S1 $\subset$ S2 or S2 $\subset$ S1

b) S1 and S2 are such that either (i) if they share more than one node, then the shared nodes also form a simple subgraph or (ii) they share at most one node and that the shared node is such that it is an entry node for one of S1 and S2 and an exit node in the other one.

*Proof*: Assume that $S_1$ and $S_2$ are simple subgraphs and that S'= $S_1 \cap S_2$. We note that when S' = $\varnothing$, i.e., when | S' | = 0, clearly $S_1$ and $S_2$ must be mutually exclusive by definition.

When S'= $S_1 \cap S_2$ = { s' }, i.e., when | S' | = 1: As $S_1$ and $S_2$ are simple subgraphs and all nodes in $S_1$ are connected, then there must be an edge between s' and $s_1$, for some $s_1 \in S_1$, $s_1 \notin$ S', and there must be an edge between s' and $s_2$, for some $s_2 \in S_2$, $s_2 \notin$ S'. As a consequence of the above and the fact that that $S_1$ and $S_2$ are simple subgraphs, s' must be the *entry* node for one of $S_1$ and $S_2$ and it must be the *exit* node for the other of $S_1$ and $S_2$.

Assume that S'= $S_1 \cap S_2$ = n > 1, where $S_1$ has the *entry* and *exit* node being $s_{en1}$ and $s_{ex1}$, and where $S_2$ has the *entry* and *exit* nodes being $s_{en2}$ and $s_{ex2}$. We show that it must be the case that the entry and exit nodes of both S1 and S2 are shared in S': { $s_{en1}$, $s_{ex1}$, $s_{en2}$, $s_{ex2}$ } $\subset$ S', where S' = $S_1 \cap S_2$.

Consider S', S' = $S_1 \cap S_2$. Clearly there must be an edge between some nodes $s_1$ and $s_2$, $s_1$, $s_2 \in$ S', as otherwise S' would not be connected and neither would be $S_1$ and $S_2$. Consider the edge between $s_1$ and $s_2$, it is either $(s_1, s_2)$ or $(s_2, s_1)$. If it is $(s_1, s_2)$, then $s_1$ is the e*xit* node of $S_1$ and $s_2$ is the *entry* node of $S_2$. If it is $(s_2, s_1)$, then $s_2$ is the e*xit* node of $S_2$ and $s_1$ is the *entry* node of $S_1$.

We showed that for S', S' = S1 $\cap$ S2:

- When | S' | = 1 the theorem is true.

- When | S' | > 1, we showed S', S' = $S_1 \cap S_2$, must be such that entry and exit nodes of $S_1$ and $S_2$ must be in S': { $s_{en1}$, $s_{ex1}$, $s_{en2}$, $s_{ex2}$ } $\subset$ S'.

We need to show that S', S'= $S_1 \cap S_2$, is a simple graph. For that, consider any node s $\in$ S'. There are only two possibilities in that (a) s is one of entry or exit nodes of $S_1$ or $S_2$ or (b) it is not.

(a) Node s ∈ S' is not an entry or exit node, i.e., s ∉ { $s_{en1}$, $s_{ex1}$, $s_{en2}$, $s_{ex2}$ }. In such a case, assume that s ∈ $S_1$. As s is neither an entry nor exit node of $S_1$, s is connected to another node of S', as otherwise s would be an entry or an exit node for one of $S_1$ or $S_2$. Thus, the node s satisfies the properties of a simple subgraph. The case for when s ∈ $S_2$ is similar.

(b) Node s, s ∈ S', is one of the entry or exit nodes of $S_1$ or $S_2$, i.e., s ∈ { $s_{en1}$, $s_{ex1}$, $s_{en2}$, $s_{ex2}$ }:

Assume that s = $s_{en1}$, i.e., it is an entry node of $S_1$. As a simple graph has only one entry node, s is connected to (i) either some $s_1$ ∈ ($S_1$ − S'), i.e., in $S_1$ but not in S', or (ii) it is connected to some $s_2$ ∈ (S' − $S_1$):

    i.   $s_{en1}$ and $s_1$ ∈ $S_1$ and $s_1$ ∉ S': ($s_1$, s) is an incoming edge from $S_1$ to S' as s is the *entry* node for $S_1$. But $S_1$ must be connected to $S_2$, which means that $s_{ex1}$ must be connected to $S_2$ via entry node of $S_2$, node $s_{en2}$. Furthermore, S' must be connected to $S_2$ by an outgoing edge ($s_{ex2}$, $s_2$) to some node $s_2$ ∈ $S_2$', where S2' = ( $S_2$ − S'). S' thus satisfies the simple subgraph properties.

    ii.   $s_{en1}$ and $s_2$ ∈ (S' ∩ $S_2$) and the edge ($s_{en1}$, $s_2$) connects $S_1$ with $S_2$. But as $s_2$ ∈ $S_2$, $s_2$ = $s_{ex2}$ as otherwise $S_2$, which is a simple graph would have two entry nodes. S' thus satisfies the simple subgraph properties.

Arguments for the remainder of cases when s, s ∈ S', is one of the entry or exit nodes of S1 or S2, i.e., s ∈ { $s_{en1}$, $s_{ex1}$, $s_{en2}$, $s_{ex2}$ }, is similar.

It should be kept in mind that Algorithm 4 has been described for ease of presentation and more efficient algorithms may exist. Furthermore, the algorithm is used only in the design process, which is a one-time task, rather than executing a smart contract once it is deployed. The final graph representing processing simple subgraphs off-chain with HSM nodes is transformed into a smart contract discussed in the following chapter. We conclude this chapter with two examples of applying the algorithm on two use-cases.

## 4.3  EXAMPLES OF ALGORITHMS UTILIZATION

### 4.3.1  Example of Seller and Buyer Escrow Deposit

The first case involves a seller and a buyer of some product that includes an escrow deposit (Asgaonkar & Krishnamachar, 2019).  After the product is posted for sale, the buyer and seller negotiate the price. Once an agreement is reached on the price, the contract is prepared and signed that stipulates escrow deposits and delivery.  Once each of the buyer and seller makes a deposit to an escrow account, then shipment, which includes crossing borders and hence involving customs, may occur that involves delivery to a port, going through customs, storing on a ship, then processed at the destination customs, unload from ship to port, buyer pickup, and finally return of escrow deposit.

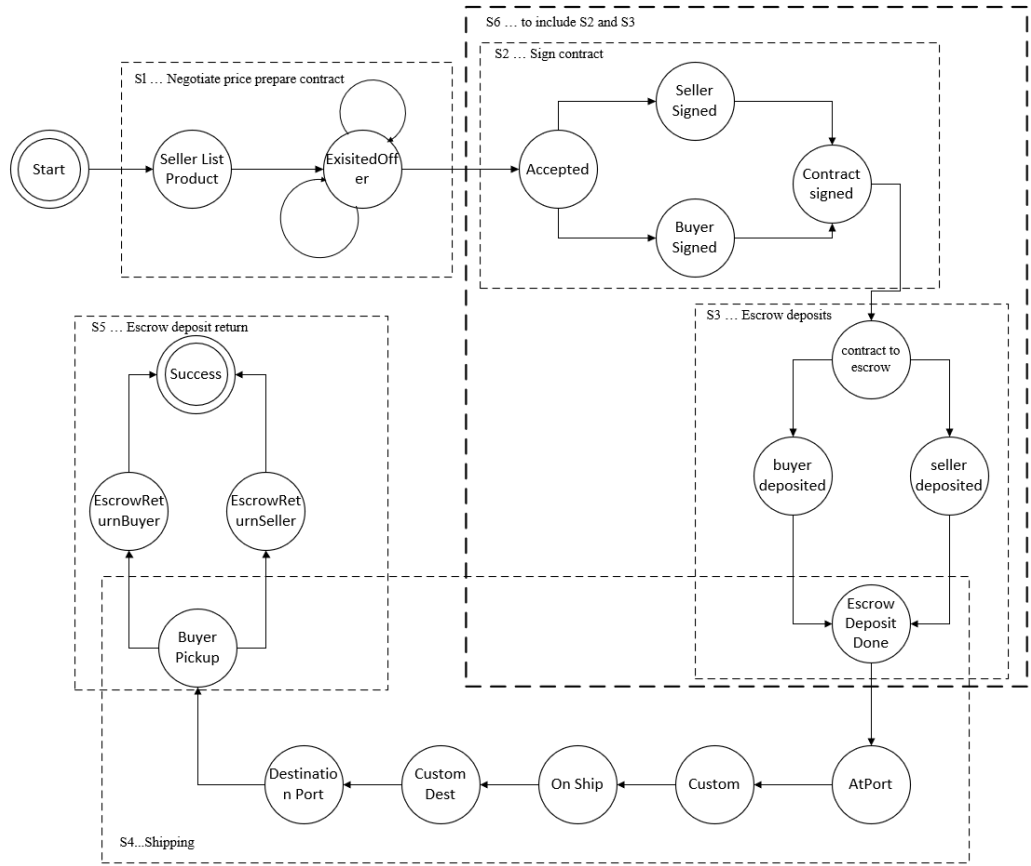

Fig.  4.5    FSM graph for buyer-seller interaction with escrow account in use

We created the FSM for this case and then applied Algorithm 4 to find all simple subgraphs.  Fig. 4.5 shows the FSM model's graph and simple subgraphs that were identified by the algorithm.

S1 … *Negotiate Price Prepare Contract* contains states representing negotiation to reach a price between the buyer and a seller.  Suppose semantics indicate that details on negotiations are not necessary. In that case, the computation can be performed off-chain as long as upon return from off-chain processing,  the negotiated price is signed by each party, confirming the agreement on the price, and is recorded on the blockchain.

S4 … *Shipping* represents a sequence of events involving distinct actors in the shipment. If processed off-chain, digitally signed information about each of the states' outcomes should be recorded on the blockchain.

*Two Parties Agreement:*

S2 … *Sign Contract* represents the signing of the contract that details the price and shipment delivery. Upon return from off-chain processing, the contract that is signed by both the seller and buyer needs to be recorded on the blockchain.

S3 …*Escrow Deposit* represents escrow deposits made by each of the seller and the buyer.  Upon return, confirmation of both deposits from the escrow account holder must be recorded on the blockchain.

S5 … *Escrow Deposit Return* represents the return of escrow deposits.  If processed off-chain, digitally signing receipt about each of the funds' return should be recorded on the blockchain.

It should be noted that two simple subgraphs, such that one is not a subset of the other, may be such that they share at most one node and, furthermore, that the shared node is an entry node for one of the subgraphs and an exit node in the other.  For instance, simple subgraphs S3 and S4 share the node Escrow Deposits Done.

There are additional simple subgraphs that were found that are not shown in Fig. 4.5, simple subgraphs that contained other simple subgraphs, such as S6 containing subgraphs S2 and S3, and others.

### 4.3.2   Example: South African Real Estate Transaction

Consider another example that involves real estate transactions over the blockchain

(Tilbury & Rey, 2019). The ESC, the proposed application name given by (Tilbury & Rey, 2019), starts with listing a property that stays listed until the buyer makes an offer received by the seller's broker. The broker prepares a purchase contract and sends it to all actors and waits for their agreements to sign the contract. Next, the agreement is signed by the seller and the buyer after verification. Next, payment is made, followed by the deed transfer, and finally all documents are uploaded to the blockchain. Generated documents include Property, Deed, Identity and Escrow. The payment may be made via cryptocurrency or through a bank. Once the seller makes payment, either via bank or cryptocurrency, payment is verified by a notary, title and deed are processed and all is recorded on the blockchain. Fig. 4.6 represents the state diagram and the simple subgraphs found by using our algorithm.

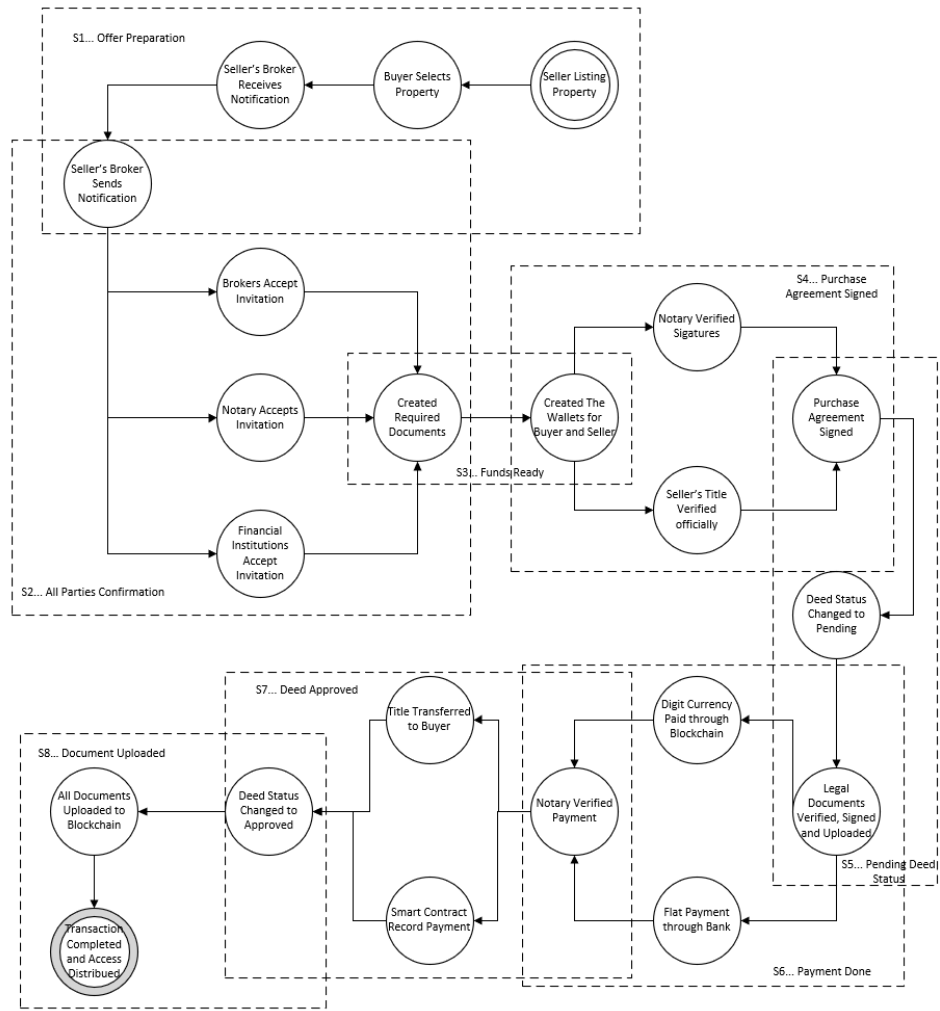

Fig. 4.6 FSM graph for Real Estate Transaction on Blockchain

The simple subgraphs that are shown are now discussed:

*The Sequence of Events:*

S1… Offer Preparation is a sequence of states starting with Seller Listing Property, then ends up with the state Seller's Broker Sends Notification.

S3… The subgraph *Funds Ready* starts with the state *Created Required Documents*, then ends up with state *Created the Wallets for Buyer and Seller* to be ready for transferring funds.

S5… Pending Deed Status is the subgraph starting with Purchase Agreement Signed, then through states Legal Documents Verified, Signed and Uploaded.

S8… *Document Updates* is the sequence of events to upload documents, wrap up the transaction and distribute the document access over blockchain at the end of the transaction.

*Two Parties Agreement:*

S4… Purchase Agreement Signed is the subgraph requiring the two parties' confirmation represents states Notary Verifies Signatures and Seller's Title Verified Officially.

S7… Deed Approval subgraph needs two parties' confirmation on both states Title Transformed to Buyer and Smart Contract Records Payment.

*Any of N:*

S6… *Payment Done* subgraph demonstrates the scenario that only one of the options needs to be done. That is, buyer can either make *Flat Payments through the Bank* or *Transfer Digit Funds* to the *Seller via Blockchain wallets*.

*M out of N:*

S2…*All Parties Confirmation* is the subgraph that ensures that all involved parties must be informed about the offer from the buyer and confirm it.

We notice that some simple subgraphs contain only a few states. For instance, *S3…Funds Ready* contains only two states*: Create Required Documents* and *Create Wallets for Buyer & Seller*. Clearly, the tradeoff between the overhead of processing off-chain and on-chain computation cost saving must be represented by applying some cost estimation model to ensure that the overhead due to off-chain processing of the subgraph is lower than processing it on the blockchain – topic of the following chapter.

### 4.3.3 Smart Blockchain Badges for Data Science Education

Mikroyannidis (2018) proposed an application called Blockchain Badges for Data Science Education based on Blockchain. The proposed application intends to integrate course offering and job hunting, and it offers badges as rewards. The general use case is shown in Fig. 4.7. The course providers and employers register on the application as the product merchants. The customers register with the system for one of the three purposes: personal interests, internal promotion, and skill training for a new job. The customers select their courses, pay money to the merchants through the platform, and then finish the required courses. Eventually, they are rewarded with badges to recognize their accomplishments.

Fig. 4.7 H-FSM Graphs as Blockchain Badges for Data Science Education

We can identify the following patterns by testifying the FSM with the algorithms defined in above.

*Sequence of Events:*

S2… *Customer Logon*:  The simple subgraph proceeds through states *System Associate Courses*, *Job Requirements,* and *Customer Registers on System*. It also  shares its *entry node* with the simple subgraph S1, where the shared node plays the rode of an *exit node*. S2 also shares one its *exit node* with S3 where the shared node is an *entry node*

S4… *Badges Rewarding* sequence of events: The simple subgraph starts with the state *Customer Register Required Courses* and end with the state *Customer Gains Badges*

S5… *Course Provider Offers Courses*: The simple subgraph starts with *Course Provider Registers on System* and proceeds, via Course Provider List Courses to the state *System Associate Courses* to *Job Requirements*. Besides, it is a proper subgraph of S1

S6… *Employer Posts New Jobs*: The simple subgraph starts with *Employer Register on System* to the state *Employer Posts Jobs*. Besides, it is a proper subgraph of S1

S7… Customer Register Courses for Job: The simple subgraph starts with Customer Seeks Jobs to the state Customer Register Required Courses. Besides, it is a proper subgraph of S3

*Two Parties Agreement:*

S1… *System Courses Preparation*: This simple subgraph requires two parties' involvements: one is *course provider registration* represented by the subgraph S5; another is *employer registration* represented by a subgraph S6. Besides, it is a superset of subgraphs S5 and S6

*Any of N:*

S3… *Customer Courses Registration*: Overlapped with S2 and S4; Parent of subgraph S7

The identified subgraphs relationships are complicated. According to Theorem 1, there are three different relationships: (i) isolation: for instance, S2 is independent of S4; (ii) Ownership: for instance, S3 completely contains S7; (iii) Overlapping: for instance, S3 shares one state with S4 – as per Theorem 1. Once the simple subgraphs are found, then we need to decide which are suitable for processing off-chain.

## 4.4  COMMENTS

Although we described the application of the algorithm on two use cases in Section 4.2 as examples, we verified the results by running the algorithm on many use-cases and the verified that Algorithm 4 found all simple-subgraphs successfully.  We also validated the found simple-subgraphs by carefully examining the semantics and the FSM model of the application and determined that the Algorithm does find simple-subgraphs that have localized goal attained by the simple subgraph when transitioning from the subgraph's exit node.

# CHAPTER 5    INTERFACE BETWEEN ON AND OFF-CHAIN PROCESSING

The previous chapter discussed the issue of What to off-chain. In this chapter, we are answering How to achieve interoperability between on and off-chain processing by the interface between on and off-chain processing. We firstly discuss the requirements for the interface and then we illustrate how to prepare the interface automatically. Next, we discussed the deployment architecture for both public and private blockchain networks. We also describe how off-chain storage may be supported.

## 5.1   REQUIREMENTS FOR INTERFACE BETWEEN ON AND OFF-CHAIN PROCESSING

In (Bodorik, et al. 2021), we stated that once a simple-subgraph pattern is slated for off-chaining, it is replaced in the FSM model with a node representing an HSM that includes the simple subgraph slated for off-chaining.  The result is an (i) HSM, derived from the original FSM model, in which simple subgraphs to be executed off-chain are replaced with HSM nodes, and (ii) a list of patterns, each represented by a simple subgraph, that are to be executed off-chain.  The HSM is transformed into a smart contract using a method described in (Marvidou & Lazska, 2018) that we augment to provide interface between the blockchain and off-blockchain computation. Recall that the transformation method uses security patterns that automatically patch security holes instead of relying on the developer to facilitate them and thus, potentially introduce bugs.

For each of the HSM nodes, representing computation that has been determined by the developer to be processed off-chain, interface is provided automatically for interaction between off-chain and on-chain computation.  Using information on parameters that need to be passed to off-chain computation, a method is provided as a part of the smart contract, to retrieve them and pass them to off-chain computation.  Another method is created that is invoked by off-chain computation when it completes its off-chain execution in order to pass the results of off-chain computation back to to be recorded on the blockchain.

Due to the properties of simple subgraphs, once there is a transition into the subgraph's entry node/state, computation is entirely within the subgraph until there is a transition into an exit node/state. That means that once the entry node/state is reached, any further computation needs to be performed off-chain until the subgraph execution off-chain reaches a state transition into the exit node/state, at which point any further method invocations need to be executed on chain. This subgraph property is used to prepare an interface between on-chain and off-chain computation automatically.

We also note that the transformation of an FSM model into a smart contract, as described in (Mavridou & Laszka, 2018), is achieved by generating smart contract methods from the states' FSM description transitions with their inputs and outputs. Each smart contract method has parameters that are defined by FSM inputs to a transition, and it returns some values/object that corresponds to the transition outputs. However, the smart contract method also reads and writes the blockchain, and that reading and writing of blockchain are also represented through transitions inputs and outputs. Thus, off-chain execution of smart contract methods may also require reading from and writing to the blockchain. For instance, a smart contract method needs to know the state of execution and also the transition to the next state upon the end of a method's execution, i.e., at the minimum, it needs to read and write the smart contracts "global" variable representing the state of computation, which we shall refer to as *state*.

However, off-chain execution cannot not access read from the blockchain unless it uses a smart contract getter method executing on chain to read data stored on the blockchain. The off-chain processing cannot write to the blockchain until its execution off-chain completes. Consequently, when transitioning to off-chain execution of a smart contract method, in addition to the parameters passed to the method, we also need to find, retrieve, and deliver to the off-chain processing any blockchain data that the method reads. Similarly, upon moving back to on-chain execution, if the method executes any statements that write to the blockchain while executing off-chain, such writes have to be recorded and handed-off, together with transition outputs, for processing and recording on the blockchain. However, before the blockchain is updated with the new state and data and returned to the on-chain execution, the off-chain execution results must be reviewed and approved by each of the participants affected by the off-chain computation

(Bodorik, et al. 2021).

## 5.2   On Chain To Off-Chain Transition

We have the FSM definition and its graph representation and also the definition of the simple subgraph S' of which processing is to be moved off-chain in.  The interface that is provided automatically is shown graphically in Fig. 5.1.  It shows two of the smart contract methods, *method1* and *method2*, and the state variable, called *state*, representing the state of the FSM computation.  Marvidou & Lazska's (2018) created a method to transform FSM definition to the smart contract.

Fig 5.1(b) outlines amendments we make to the smart contract methods to facilitate the interface between on and off-chain computation. We introduce a new state variable in a smart contract, a variable called *offChain* that is stored on the blockchain and indicates if the contract method invocation should be executed on or off the chain.

Assume that the simple subgraph has an entry node/state $s_{en}$ and an exit node/state $s_{ex}$. The following description applies to each of the smart contract methods that is to be executed off-chain.  Although the state variables *state* and *offChain* are global, the events, *offChainEvent* and *offChainDoneEvent*, are unique to each method and hence are shown in figures as *offChainEvent1* and *offChainEvent2* and *offChainDoneEvent1* and *offChainDoneEvent2*.

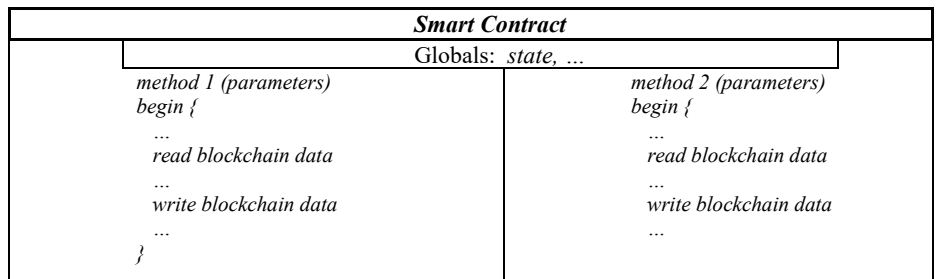

(a)   Smart Contract and its methods

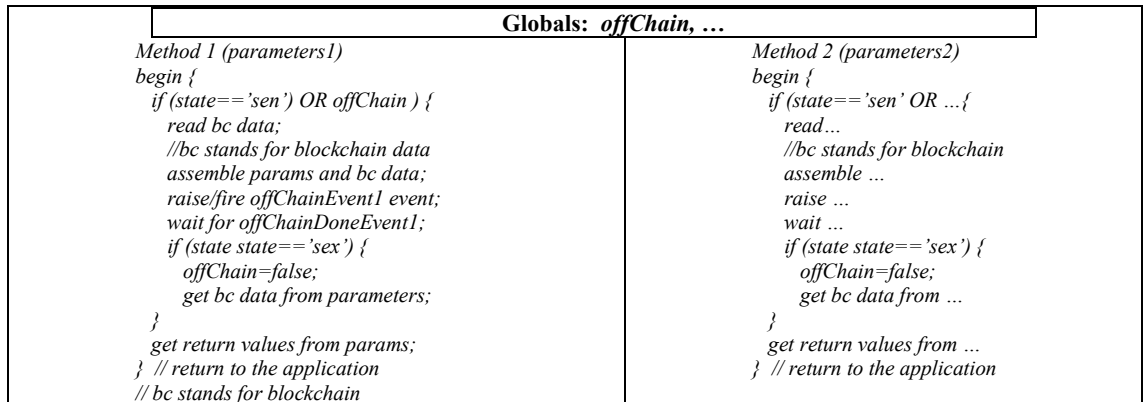

(b)Modifications for on-chain to off-chain interface

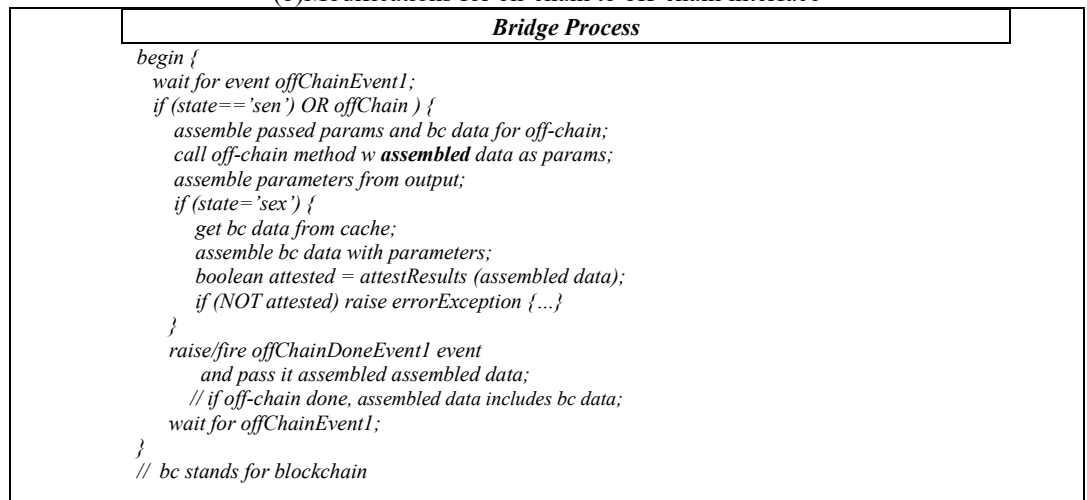

(c) Bridge1 process activities

Fig. 5.1 Interface Between On-chain and Off-chain Processing

Each of the smart contract methods is checked and when there is code to transition to the subgraph's *entry* node, $s_{en}$, the variable *offChain* is set to true to indicate that any further method invocations should be to a method performed off-chain. Furthermore, if *offChain* is true when a method is invoked or the state of computation reached the transition into the state $s_{en}$, i.e., when state == ' $s_{en}$ ', we insert the code to raise an event and pass it parameters that include:

Identification of the smart contract method

Parameters of the smart contract method

Smart contract values/objects read from the blockchain, i.e., the values and objects that are retrieved from the blockchain and are referred to as blockchain data.

Invocation of off-chain methods once off-chain computation starts

Once the parameters and blockchain data are assembled, they are passed as parameters to the *offChainEvent* that is fired/raised and for which there is a corresponding process that is waiting on the event and that will instrument/call the off-chain execution. (Actually, it is a thread of a bridge process, but for simplicity we shall refer to it as a *bridge process*.) The process that was executing the smart contract method and fired the event suspends itself waiting for an event, called *offChainDoneEvent*, that is raised when the off-chain execution of the smart contract method completes.

When the fired *offChainEvent* awakens the bridge process (see Fig. 5.1(c)), it receives parameters passed to it and assembles them into an invocation of the method executing off-chain that corresponds to the smart contract method being invoked. Thus, the off-chain method needs to be "a mirror image of the smart contract method with the exception that instead of reading and writing data on the blockchain itself, it reads blockchain data from its local cache that persists from call to call as long as there is not a transition to the exit state of the subgraph, node sex. Thus, any reading of blockchain data is retrieved from the local cache that is seeded with blockchain data passed as parameters to the *offChainEvent* event. Any data written to the blockchain is written to the local cache.

Fig. 5.1(c) shows the script outline for the bridge process. It waits for the *offChainEvent* and retrieves parameters passed to the event that contain input to the smart contract method and state variables *state* and *offChain*. In addition, they include all blockchain data read by the method, which is saved in a local cache if it is not there already. Only then is the off-chain method, which corresponds to the smart contract method that was invoked, invoked while also passing it parameters. When the off-chain method reads or writes data from/to the blockchain, it actually uses its local cache. However, should the blockchain data be required that is not be present in the cache, that data needs to be obtained by pausing the off-chain method execution and invoking an on-chain getter method to retrieve the object from the blockchain and pass it to the off-chain computation. Finally, the execution of the interrupted off-chain method may proceed.

## 5.3   OFF-CHAIN TO ON-CHAIN TRANSITION

Once the off-chain method completes but the off-chain computation is not finished (off-chain computation is finished only if $state = s_{ex}$), its output contains return values that need to be passed as parameters to the *offChainDoneEvent* that is raised and then the bridge process suspends itself waiting for another call of the smart contract method. If the *state* =/ $s_{ex}$, the process waiting on the *offChainDoneEvent* retrieves the parameters of the event and returns them to the application process that caused the invocation of the method to be executed off-chain.

Off-chain execution thus proceeds as long as there is no transition into the *exit* state, $s_{ex}$. When transition to the exit state is detected upon exit from the method executed off-chain (i.e., when $state == s_{ex}$), in addition to output parameters, it also marshals all blockchain values in its cache that were written by off-chain execution and passes the parameters to a method to check the results of computation, *attestResults* method. This method needs to inform all affected parties/actors about the computation so that the actors will run their attestation process to check the results and digitally sign their approvals (of the results). Once that is done, the results, including digital signatures signifying off-chain computation approvals, are returned and marshaled into parameters passed to the *offChainDoneEvent* that is raised/fired. The bridge process then suspends itself waiting for the next *offChainEvent* to be raised/fired.

When the process waiting for the *offChainDoneEvent* event awakens and detects that off-chain processing is finished (in Fig. 5.1(b), it is the if statement if (state *state* == $s_{ex}$) *{ ... }* ), it performs the following:

- Changes the blockchain variable offChain to false.

- Writes to the blockchain values that were written to the local blockchain cache in off-chain execution.

- Returns from the smart contract method execution with the return values.

We make a couple of notes. First, we provided only a high-level description while omitting many details that we are still addressing. For instance, although they pose significant challenges in prototyping, exceptional cases and error conditions are ignored here.

Another note is that we do not provide the code for off-chain execution as that depends on off-chain computing platform; however, we do provide the interface between on and off-chain execution as described here. We are investigating generating full methods to be executed on an off-chain platform compatible to that of the blockchain platform. For instance, when a smart contract is executing on Hyperledger platform, we are investigating how to prepare an interface and smart contract methods to be executed by an off-chain platform in a separate private channel or on a platform that is also Hyperledger but executing as a private platform separate from the main blockchain platform.

## 5.4 INTERFACE CONSIDERATIONS FOR HYPERLEDGER AND ETHEREUM

There are many blockchain network products. Overall, blockchains can be categorized into two types: public and private. A public blockchain is a non-restrictive, permission-less distributed ledger system. For instance, Bitcoin and Ethereum are the most popular public blockchain networks (Data-flair.training, 2018). A private blockchain is a restrictive or permission blockchain operating in a closed network, such as Hyperledger. Although both public and private blockchain networks share the same the proposed off-chain modeling concepts, physically there are differences between them in terms of how the off-chain model interface facilitates communication between on-chain and off-chain. We explain these differences in the following subsections.

### 5.4.1 Interface for Ethereum

Off-chain communication mechanisms are still evolving for Ethereum. At the time of this writing, example for off-chain mechanisms available for deployment include state channels, Plasma, and sidechain. We focus on a sidechain solution because it can scale the general-purpose application and secure blockchain's natural characteristics by running on a compatible EVM. xDai is an example of such a sidechain (xDai Chain Network, 2020). A typical architecture for deployment is depicted in Fig. 5.2.

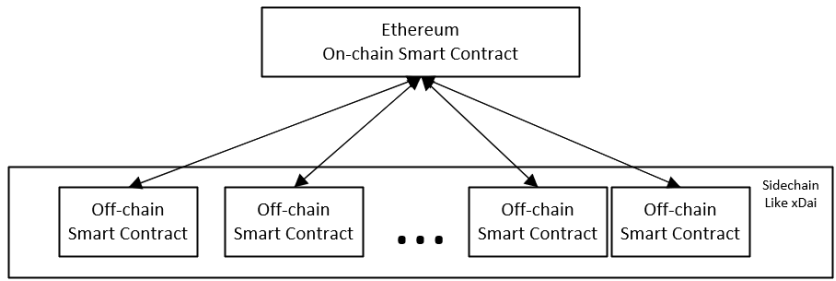

Fig. 5.2    Ethereum Network on/off-chain data exchange

Traditionally, the communication between sidechain and Ethereum network, referred to as mainchain, relies on the application layer, such as web3 API, to allocate an address for a smart contract and provide for messaging between the mainchain and sidechains. However, this mechanism compromises the correctness and trust on the blockchain because the messages transferred through an application layer could be altered accidentally or maliciously without any knowledge by the mainchain or sidechain networks. Therefore, currently emerging technologies support interaction between smart contracts deployed on different chains: sidechain contracts can invoke mainchain contracts and vice-versus.  And all this is officially supported by the interfaces provided by Oracle, such as *Chainlink*, on both mainchain and sidechain/off-chain ((Markus, 2020), (Robinson, 2020)).

Fig. 5.2 depicts the straightforward architecture of data exchange between on/off-chain network. Mainchain has the deployed smart contract, while sidechains would have patterns slated for off-chain execution.

Equipped with up-to-date sidechain techniques, the smart contract on the mainchain can directly invoke the off-chain event when necessary. And the result of computation on sidechains can be submitted relatively easily to mainchain at the end of the off-chain computation. Besides, since the sidechain is another Ethereum compatible blockchain network, it inherits all the characteristics of Ethereum, including immutability, transparency, security and decentralization. Therefore, all participants of the transaction can also check the "*off-chain*" computation on sidechain.

### 5.4.2   Interface for Hyperledger

Hyperledger Fabric, on the other hand, provides a collaborative platform for implementing Member service, consensus, and Smart contract (chaincode) (Hyperledger Fabric, 2018). Note that in Hyperledger, the terms of smart contract and chaincode differ slightly from those of other blockchains. A smart contract in Hyperledger is a struct type of actors, while chaincode provides methods to be invoked by applications. However, since they work together as a whole, we call them together as smart contracts to make the concept consistent with other blockchains. Smart contracts are defined globally but initialized inside channels (Hyperledger Fabric, 2018). Therefore, they can be invoked internally in the same channel or externally across channels. All channels share the data storage called "world state." At the time of writing, the chaincode (smart contract) can be implemented in three languages, namely JavaScript, Go, and Java (Pattnaik, 2020). As an aside, JavaScript is becoming more and more popular because it is expressive and relatively easy to use. Furthermore, Hyperledger fabric provides easy access to off-chain DBs, such as couch DB (Pattnaik, 2020). The concept of Hyperledger smart contracts' data exchange is graphically shown in Fig. 5.3.

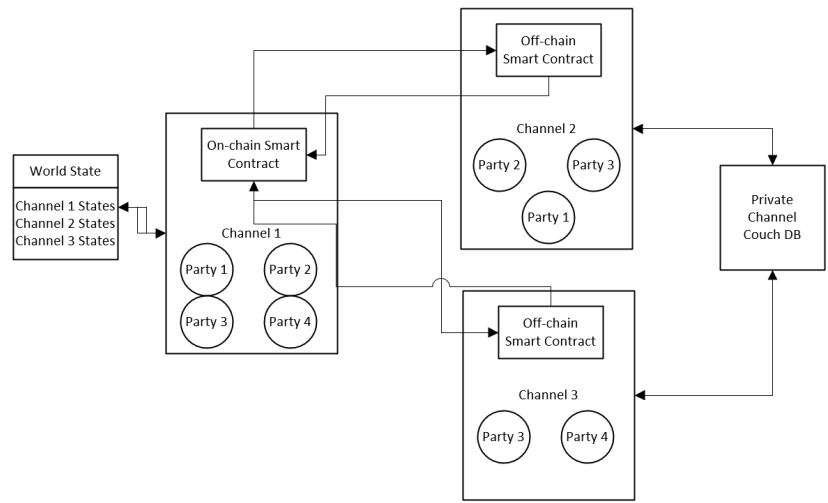

Fig. 5.3    Hyperledger Network on/off-chain data exchange

As mentioned above, smart contracts are initialized in channels. Each channel has a predefined access list of the authorized members. Each channel can be treated as one relatively independent blockchain, while the channels are able to exchange data wherein access control to the data is configured independently for each channel. Fig. 5.3 depicts a typical use case of how smart contracts exchange data on Hyperledger with our proposed modeling approach. We treat Channel 1 as the mainchain that contains the so-called "on-chain" smart contract and interfaces of interoperating "off-chain" smart contracts for off-chain computation, while the "off-chain" smart contracts are running over channels 2 and 3. Channel 2 is configured to be accessed by party 1, 2 and 3, who are the participants involved in the pattens processing to the off-chain. Similarly, channel 3 is configured for party 3 and 4 only, which maps to the participants of the off-chain pattern. Both channel 2 and 3 can access the off-chain database, such as CouchDB, running on the same Fabric framework with Hyperledger.

More importantly, Channel 1, which is treated as mainchain, can update the global state variables that are ready-only for the channel 2 and 3. For instance, channel 2 keeps watching the global variable which triggers its off-chain event. Once the off-chain event is triggered, Channel 1 starts to watch the global variable, indicating the corresponding off-chain event's status. As this variable indicates the off-chain event is done, channel 1 starts to attest the computation results of channel 2 and then updates the state variables.

## 5.5   OFF-CHAIN STORAGE

Eberhart (2018) stated that off-chaining computation requires data persistence mechanism whenever communication happens between on-chain and off-chain computation. Fortunately, both the Ethereum sidechains and Hyperledger provide sophisticated technology to access off-chain databases, such as *LevelDB* for Ethereum and *CouchDB* for Hyperledger. A No-SQL database is preferred for both structured and unstructured data if large volume data is involved (Karafiloski & Mishev, 2017). Our work focuses on what data structure(s) should be persisted/updated in the storages on both on-chain and off-chain in terms of interoperability rather than from the perspective of security.

For instance, Ethereum clients may use two different database software solutions to store their tries: (i) *Rocksdb* is built over *LevelDB* and is scalable to run on servers with many CPU cores to efficiently use fast storage to support IO-bound, in-memory, and write-once workloads and to be flexible to allow for innovation. (ii) *LevelDB* is a commonly used data storage for system logging that provides platform persistence. It is an open-source Google key-value storage library that offers forward and backward iterations over sorted maps  from string keys to string values, custom comparison functions, and automatic compression. The state *trie*, referred to as Merkle Tree of Ethereum, provides a key-value pair for every system account. The "key" is a single 160-bit identifier, which refers to the address of the Ethereum account. The "value" in the state *trie* is generated by encoding the Ethereum account's account details. The state *trie's* root node, which stands for a hash of the current state *trie*, can be used to identify the state trie; the state *trie's* root node cryptographically depends upon all internal *state-trie* data (Applicature Team, 2018).

We identified two approaches to persist updated data on the blockchain. One is utilizing the "*extraData*" field *trie*, while the other is emitting a customized event message in the "data" field of the log table, which can be viewed in real-time on Google *BigQuery* (Google BigQuery, 2019). The purpose of off-chain storage is two-fold: (i) attesting the results from off-chaining computation; (ii) storing the data generated by transactions and associate the data with its hash code. The data structure we defined for off-chaining computation records stored in the off-chain database is shown in Fig. 5.4.

| Data Structure of Off-chaining Computation | |
|---|---|
| Key: Value | |
| TransactionID | Byte[] - It indicates the current transaction identification. |
| CurrentOffchainPatternAsSubmitted | Byte[] – It indicates which off-chain computation invocated by this remote call. |
| SmartContractFromAddress | Varchar – The physical address of Smart Contract Byte[] From. |
| SmartContractToAddress | Byte[] – The physical address of Smart Contract Calling to. |
| ParticipatedActors | Enum – The list of participants identifications that are defined universally. |
| ParticipatedActorsSignatures | Enum – The list of participants' signatures. |
| TransitionParameters | Enum – The list of inputs or outputs of the selected subgraph. |

Fig. 5.4   Data Structure for Off-chaining Computation

It should be noted that a transaction, in the context of blockchains generally means a set of values written to the blockchain on Ethereum and the to the state world DB in Hyperledger, by one method. Each write is time-stamped and the blockchain mechanism ensures that the updates occur in the order as defined by their timestamps. In addition, Ethereum also automatically charges, the account executing the method issuing writes to the blockchain, for gas to execute that method's instructions and also for storage.

Key-Value data structure defined above should store the following information about transactions on both on-chain and off-chain storage: [*TransactionID, CurrentOffchainPatternAsSubmitted, SmartContractFromAddress,*

*SmartContractToAddress, ParticipatedActors, ParticipatedActorsSignatures, TransitionParameters(updated data)*]. More specifically, *TransactionID* is required when multiple transactions are executed simultaneously. *CurrentOffchainPatternAsSubmitted* identifies the off-chain pattern. *SmartContractFromAddress* and *SmartContractToAddress* are useful when developers want to troubleshoot issues arising with the transaction. They can also be used to verify and validate submitted inputs and outputs parameters from authorized resources. *ParticipatedActors* is used to record involved participants in the current computation. *ParticipatedActorsSignatures* contains the list of signatures of engaged participants. *TransitionParameters* contain actual updated data. With the structure defined above, we can satisfy the two major functionalities, *persist* and *verify*, used for the interoperation of on-chain and off-chain computation.

Similarly, the Hyperledger private blockchain network can easily access off-chain databases, such as *LevelDB* and *CouchDB*, running on the same platform as Hyperledger. Therefore, the data structure used for records that persist on both on-chain and off-chain storage is identical to the Ethereum Public Blockchain network's data structure.

## 5.6 SUMMARY

In this chapter, we addressed the problem of how to process the pattern off-chain, that is, we investigated the protocol for communication between on and off-chain processing in terms of messages that need to be exchanged and their content. We also developed a data structure to support the interface.

# CHAPTER 6 DECIDING TO PROCESS A PATTERN OFF-CHAIN

The previous chapters tackled *What* and *How* to process off-chain while, in this chapter, we are answering *When* to process the off-chain. In Chapter 4, we built an algorithm that analyses the FSM graph defined by the set of states' *S* to find patterns to be considered for processing off-chain. However, not all patterns that are found by the algorithm are suitable for processing off-chain – the interaction between on and off-chain processing incurs overhead that must be offset by the benefits of processing off-chain. A simple-subgraph pattern may not be suitable to be processed off-chain due to the application's semantics that is not considered by the algorithm. Thus, the developer needs to examine a simple-subgraph pattern in terms of details on inputs and outputs, semantic information, if available, and expected overhead cost so that she/he may decide whether to process a design pattern off-chain. However, before we proceed, we review the blockchain properties that should be maintained even if some of the processing is off-chain.

*Immutability*: Using hashes preserves immutability. Data held off-chain is signed on the blockchain with a hash code. If data is stored off-chain, the hash code of data, retrieved from off-chain storage, is recalculated and compared to the data's hash code held on the blockchain to ensure that it has not been modified while stored off-chain.

*Availability*: As the availability of its components determines the system availability, whether or not and how the availability of off-chain processing/storage is affected depends on the off-chaining implementation. For instance, if the off-chain computation is handled by a sophisticated public sidechain network, such as xDai, the off-chain computation's availability should not be an issue (xDai Chain Network, 2020). Industries had proved the reliability of several well-known sidechain networks because the sidechain products can provide persistent network connections with end-users. On the other hand, the availability of private blockchain networks like Hyperledger relies on the host servers' performance and the host servers are managed by the blockchain-network participants. Fortunately, current popular public cloud providers support private Blockchain services and thus the availability of private blockchain networks can benefit from cloud services.

_Trust_:  Maintaining the same trust on both on-chain and off-chain computation is a challenge.  Blockchain and its smart contracts secure trust as anyone may analyze the smart contract and examine the state variables (in permissioned blockchains, as long as they have access rights).  In fact, the blockchain is replicated and stored on a network node and, furthermore, each node performs the same smart contract computation --  the results are considered valid only if the majority of nodes agree on their validity.  The blockchain's trust may be negatively affected when parts of the computation are moved to off-chain processing because the participants no longer may be assured that the script moved off-chain has not been altered and hence is not producing false results.

It should be noted that, conceptually, the off-chain computation may be shown to be trustful by ensuring that it (off-chain computation) returns information on all inputs to transitions, transitions' computation and their outputs.  That information may be stored on the main blockchain and then examined for its correctness that, in the general case, would be equivalent to re-computation of the same results on the main chain and thus incurring additional overhead costs that is likely to exceed any savings made by off-chain processing.

## 6.1  DECISION-MAKING DATA REQUIREMENTS

The developer needs to consider the semantics of computation and decide which data generated through the off-chain execution needs to be passed to the blockchain upon completion of the blockchain execution, such that it is sufficient for affected parties to assure themselves that the off-chain computation is correct.  A complementary issue is how an actor, affected by off-chain execution, can verify the results and attest they are correct so that they can be recorded on the blockchain.  Clearly, guidance to the smart contract developers in this aspect is required.  More specifically, the smart contract developer needs to consider and determine:

- Information to be obtained/retrieved from blockchain and provided to off-chain computation. Such information is obtained mostly from inputs into transitions, including information transmitted to the smart contract methods as an input to a state transition.  As script statements of a processed off-chain pattern do not have direct access to the blockchain data, any data from the blockchain that may be required for computation must be supplied as input at the beginning of a processed off-chain pattern or it has to obtained by off-chain computation by invoking the smart contract getter methods to retrieve such data from the blockchain. Similarly, any writes made to the blockchain must be collected and, upon return from off-chain processing to on-chain processing, must be returned to the continued on-chain computation so that they can be recorded on-chain. However, for some off-chain patterns, such as the escrow deposit two-parties pattern, we cannot prepare all the required data from the main chain at the beginning of the off-chain computation. The escrow deposit pattern requires information on the availability of sufficient funds by both parties – and the availability of funds is stored on the main chain as an account balance. In this case, we should not process such a simple-subgraph pattern off-chain because its overhead cost due to off-chain computation is likely to be too high to justify off-chain processing, particularly when the off-chain processing is simple and requires data to be retrieved from the mainchain.

- Information to be provided to the blockchain for recording after the off-chain processing is completed. This information is derived from state transformations' outcomes and passed to off-chain processing. Of necessity, any writes to the blockchain made by off-chain execution must be committed to the blockchain once the off-chain computation completes.

- Information required to ensure that off-chain computation and its outcome are the same as when the computation was carried out on the blockchain. This information may be in addition to the data that needs to be returned at the end of the computation. It is derived from the smart contract semantics and may require attestation by the affected parties confirming that they agree with the computation results. Usually, attestation is confirmed by the affected parties by examining the results of computation and associated information and signing them with the actors' digital signatures that are stored, together with the results of off-chain computation, on the main chain.

- Overhead cost for interface to facilitate off-chain computation. This overhead cost

needs to be determined for each method and compared to the benefits and then summarized before it can be determined that savings will indeed be achieved. This analysis is required even if the off-chain processing cost is negligible. The interaction between off-chain and on-chain computation incurs overhead cost due to the interface between the main and off-chain execution. Further complication also arises in terms of the units being used for overhead cost computation: costs expressed in terms of time-delay may lead to tradeoffs that may differ in their properties than when costs are expressed in monetary units. For instance, in monetary considerations, the off-chain processing cost may be negligible, while it may not be deemed to be negligible if the cost is in terms of delay.

The process for decision-making whether to process a pattern off-chain can be described simply as:

The designer examines the simple-subgraph patterns and their contextual information on an iterative basis, one simple-subgraph pattern at a time, in order to decide whether or not patterns should be processed off-chain.

If the developer decides that a simple-subgraph pattern should be processed off-chain, then the proper subsets of that simple-subgraph pattern are also processed off-chain automatically. Once the decision is made, the FSM $F$ is changed to represent that decision (to process off-chain) by transforming FSM $F$ into HSM $H$ by replacing the simple subgraph in the FSM graph with a node representing the simple subgraph to be processed off-chain.

Consider, for example, the simple subgraph S5 that is in Fig.4.6 and that is repeated in Fig. 6.1(a) for the reader's convenience. The pattern represents the product shipment and, if the developer were to decide that the subgraph should be processed off-chain, then the FSM graph would be modified as shown in Fig. 6.2(b).

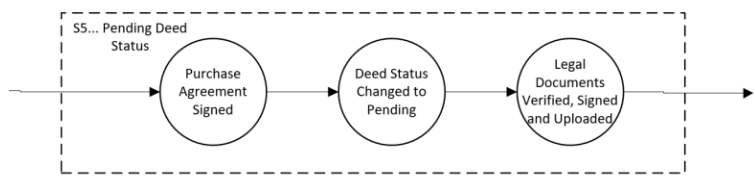

(a) Simple subgraph before the decision to off-chain

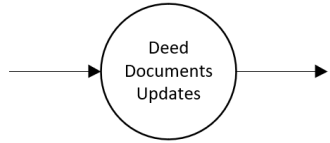

(b)  Simple subgraph after the decision to off-chain

Fig.  6.1    Escrow deposit simple subgraph before and after off-chain decision

It should be noted that the developer reviews the simple-subgraph patterns, when determining whether they should be moved off-chain, in descending order as defined by the number of other simple subgraphs that a simple subgraph has. Generally, partial results formed by processing off-chain cannot be trusted before the off-chain calculation is complete. The on-chain smart contract cannot monitor the off-chain computation process once the computation off-chain starts until the smart contract receives the results from off-chain computation and receives attestation from the parties affected by the off-chain computation that those results are correct.  Before transitioning from on-chain to off-chain computation, blockchain needs to be read to retrieve information that off-chain computation requires. Similarly, when off-chain calculation completes, before return to on-chain computation, information from transition outputs and data that needs be written to the blockchain must be compiled and provided upon return to the smart contract execution on the blockchain.  Upon return to blockchain execution, information returned by the completed off-chain execution must be examined and attested as correct before it is recorded on the blockchain.

In the previous chapter, we investigated a two-party agreement use case shown in Fig. 3.4 and repeated in Fig. 6.2 for the reader's convenience.  Fig. 6.2. illustrates a contract that is ready to be signed by two parties. It contains the states (*s1, s2, s3,* and *s4*) representing (*Contract-prepared, Signed by Party A, Signed by Party B, Contract-ready*), respectively.

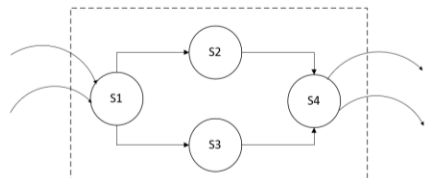

Fig.  6.2    Two-party approval

We consider simple-subgraph patterns for processing off-chain because of their property of having only one *entry* node and one *exit* node that limit the scope of activities to obtaining an agreement that certain activities were successfully completed by the parties involved. Furthermore, we are also looking for patterns that produce evidence of completion that is verifiable by all parties affected by the pattern so that attestation for correctness may be signified by digitally signing the agreement. For the two-party agreement example, successful/authorized completion of off-chain computation require the parties' signatures on the approved off-chain computation results. When applying for a mortgage, signed documents (mortgage request, insurance quote, land survey, approved mortgage) provide proof of completing the activities. Although the algorithm depicted in Fig. 4.4 identifies the simple subgraph as suitable for processing off-chain, overhead costs are likely be too high for this specific use case. Recall that we assume that storage of larger objects is off-chain regardless of whether they are produced on or off-chain.

Optimally, the size of off-chain computation results should be minimal as they need to be recorded on the blockchain. When a large amount of data is returned by off-chain computation, the size reduction efficiency is negatively affected. If it is too large, the benefits of moving a simple-subgraph pattern to off-chain may be more than off-set by the overhead. Therefore, blockchain designers should carefully examine the tradeoffs between data reduction and communication overhead between on-chain and off-chain computation at the time of deciding when to process simple subgraphs on off-chain. For that purpose, we build a cost estimation model to evaluate the cost of on-chain and off-chain computation in order to provide appropriate information when deciding whether to process a pattern off-chain.

## 6.2   COST ESTIMATION MODEL FOR ON-CHAIN TO OFF-CHAIN INTERFACE

This section develops a model for the trade-off between on and off-chain computation by estimating gained benefits vs the overhead-cost. Section 6.2.1 explains the technique we utilize for the cost estimation; section 6.2.2 describes the cost estimation model works while in the last subsection, we illustrate how to utilize our cost estimation model.

### 6.2.1   Analytical Estimation Approach of Cost Evaluation Model

Several approaches may be adopted for the cost model: Empirical, Heuristic, and Analytical methods (Hammad, 2021). The empirical approach relies on the experience gained from previous cases that are similar when moving patterns of smart contracts to off-chain. Clearly, as this research is in its nascent stage, such experience is lacking. The heuristic approach is applicable for specific scenarios in which assumptions are made about which aspects of the problem are significant and which are not – such approaches do not apply to a general model, but rather are more applicable to specific instances of a general model and, again, experience with scenarios is lacking. Analytical estimation relies on analyzing what work is required and estimating the cost of such work. In this technique, the task is divided into its basic component operations/elements for analysis. Hence, as the analytical estimation technique has scientific basis and does not draw on previous usage experiences (Hammad, 2021), we derive an analytical model based on breaking down the problem into parts for which we can create models to estimate their costs.

### 6.2.2 Cost estimation model

We build a cost model for (i) computation of a pattern on the main chain; (ii) computation for the same pattern off-chain, and (iii) overhead cost of interaction between on and off-chain computation. We develop the model for each of the parts (on vs off-chain) so that we can make the cost comparison between them that also includes the overhead cost. The evaluation process can be represented by the workflow shown in Fig. 6.3.

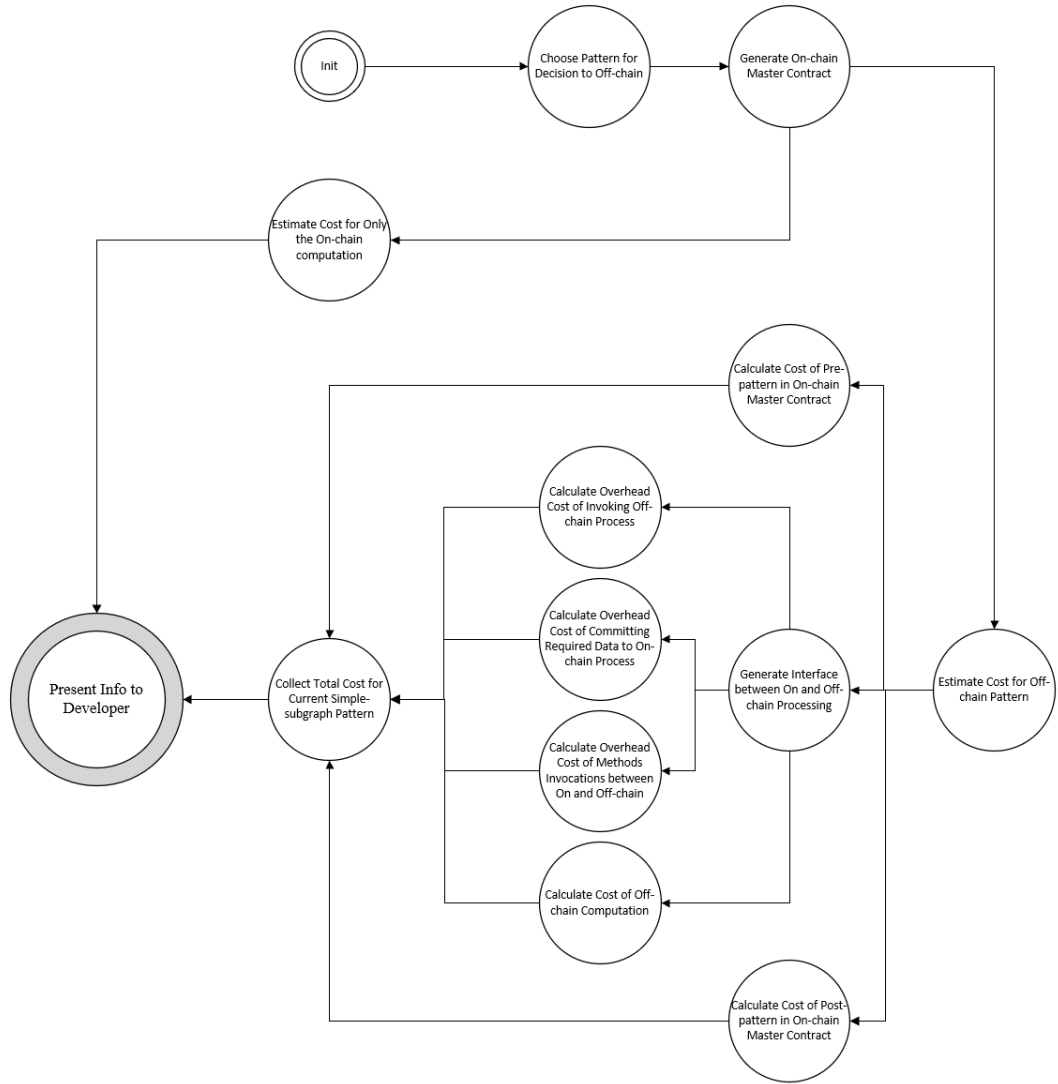

Fig. 6.3    Cost Estimation Workflow Diagram

The *Cost Estimation Workflow* diagram in Fig. 6.3 illustrates how the cost evaluation model works when deciding on one simple-subgraph pattern whether or not it should be processed of the chain. Recall that the simple-subgraph patterns are identified already by Algorithm 4 (Chapter 4) and then each pattern is examined to decide whether it should be processed off-chain. The workflow of Fig. 6.3 is used to evaluate the benefits vs overhead costs, such as delay or monetary cost, of a pattern processed off-chain. If the developer decides that a specific pattern represented by the simple subgraph S should be processed off-chain and that pattern contains any simple subgraph S' that is a subset of S, then S' will be processed of chain also as a part of S and hence S' is removed from any further consideration whether to be processed off-chain or not.

The workflow starts with selecting the pattern to be considered (recall that the patterns are considered in a descending order on the number of subgraphs pattern has). Briefly, the workflow estimates the cost of processing the whole smart contract on chain and the cost of the contract, including overhead due to off-chain processing, when the selected pattern is processed off-chain and then the two costs, together with information on the costs of various components and context information, are presented to the developer for the decision-making. We discuss further workflow details below.

Once the pattern is selected for consideration, the smart contract is generated and its cost of executing it on the main chain is estimated. To estimate the cost of the contract when the selected pattern is executed off-chain, the smart contract cost is broken down into three components:

- The cost of the contract execution on the main chain until its execution transitions to the *entry* node of the pattern (until a method of the pattern slated for off-chaining is invoked) – that part of the smart contract is referred to as pre-pattern execution.

- The cost of executing the selected pattern off-chain, such that the cost includes all overhead cost of transitioning to the off-chain, the cost of invocation of the pattern's methods that are executed off-chain, and the cost of transitioning from the off-chain back to the main chain when the pattern execution is completed.

- The cost of the contract execution on the main chain from the point when the pattern execution completes its off-chain execution, and the contract execution returns to the main until its completion – that part of the smart contract execution is referred to as post-pattern execution.

To estimate the overhead cost of off-chain processing of the pattern reasonably accurately, we calculate the overhead cost due to (i) invoking off-chain process (*Invoking Off-chain Process* on the diagram); (ii) the cost of on-chain execution to invoke the off-chain methods (*Cost of Invocations between On and Off-chain*); (iii) the cost of computation off-chain for the pattern, this step is denoted as Calculate the *Cost of Off-chain Computation*; and (iv) the cost of transitioning the computation back to the main chain once the off-chain computation completes (*Cost of Return to main chain*). The overhead cost of executing the pattern off-chain is estimated and the cost of pattern executed off-chain are added to the pre- and post-pattern execution costs to estimate the cost of the master contract execution when the selected pattern is executed off-chain.

Finally, the two costs, the cost of smart contract execution fully on-chain and the cost of the smart contract when the selected pattern is executed off-chain, are presented to the developer for her/his consideration and decision making.

### 6.2.3   Example of cost estimation on Ethereum network

In this subsection, we give an example to explain our cost estimation model further. We use Ethereum as an example to apply the cost estimation model because gas refers to the cost necessary to perform a transaction on the network on the Ethereum blockchain (Frankenfield, 2021).  Furthermore, Ethereum provides detailed gas costs for smart contract execution statements and can be used as a unit of measurement when representing actual costs.  In Ethereum, when a smart contract method is executed that writes to the blockchain, gas is actually "spent" as the gas cost of execution is subtracted from the user account that invoked the contract method. (As a note, a smart contract method that does not write to the blockchain does not cost any gas.)  In contrast, some other blockchain networks, such as Hyperledger, do not have a standard unit of measurement for computation cost estimation, and it depends on the cost of running network nodes participating in the blockchain and executing smart contracts and the cost of execution off-chain.

Regarding Ethereum, the smart contract is executed by the Ethereum Virtual Machine EVM.  An individual statement of a smart contract can be executed in Stack, Memory, or Storage. The gas consumption for these three places varies. In essence, Stack is the place where Solidity stores local simple variable values defined in functions. Memory is an area of memory on each EVM that Solidity uses to store temporary values. Values stored here are erased between function calls. And storage is where state variables defined in a smart contract reside in a smart contract data section on the blockchain (Aschenbach, 2018).  And the cost of each operation is known and is shown in Table 6.1.

Table 6.1    GAS Consumption on Stack, Memory, and Storage (Palau, 2018).

| ZONE | EVM OPCODE | GAS/WORD | GAS/KB | GAS/MB |
|------|-----------|----------|--------|--------|
| STACK | POP | 2 | 64 | 65,536 |
| | PUSHX | 3 | 96 | 98,304 |
| | DUPX | 3 | 96 | 98,304 |
| | SWAPX | 3 | 96 | 98,304 |
| MEMORY | CALLDATACOPY | 3 | 98 | 2,195,456 |
| | CODECOPY | 3 | 98 | 2,195,456 |
| | EXTCODECOPY | 3 | 98 | 2,195,456 |
| | MLOAD | 3 | 96 | 98,304 |
| | MSTORE | 3 | 98 | 2,195,456 |
| | MSTORE8 | 3 | 98 | 2,195,456 |
| STORAGE | SLOAD | 200 | 6,400 | 6,553,600 |
| | SSTORE | 20,000 | 640,000 | 655,360,000 |

We notice from Table 6.1 that the storage usage is a critical factor impacting the computation cost and hence we need to consider how much storage is used for each executable statement defined in the smart contract. Besides data usage, operation type is another consideration for the cost calculation. Table 6.2 categorizes the basic executable statements in smart contracts based on the operation types. Note that executing smart contract functions defined off-chain does not involve any cost because methods on side/off-chain are treated as read-only; in other words, side/off-chain smart contracts do not directly alter the on-chain data until the end of processing the off-chain pattern (Aschenbach, 2018) and hence do not consume gas the end of off-chain computation when data is returned to on-chain computation and is written to the main blockchain. However, the functions defined in interfaces to invoke the off-chaining computation consume gas because the parameters are passed by loading them in either memory or storage using *SLOAD* operation. Upon the completion of off-chain computation, results are returned to the on-chain computation by using *SSTORE* operation.  Table 6.2 shows further details on EVM operations.  Although we do not show the cost of gas for each low-level individual operation, such costs are known. Furthermore, tools exist for Ethereum, such as Ethereum Remix, that calculate the detailed cost of gas for executing any group of statements or a method and that is the reason why we chose Ethereum as an example.

Table 6.2   Examples of Different Operations on Blockchain (Ethereum Yellow Paper, 2021)

| Operation | Description |
|-----------|-------------|
| ADD/SUB | Arithmetic operation |
| MUL/DIV | Arithmetic operation |
| ADDMOD/MULMOD | Arithmetic operation |
| AND/OR/XOR | Bitwise logic operation |
| LT/GT/SLT/SGT/EQ | Comparison operation |
| POP | Stack operation |
| PUSH/DUP/SWAP | Stack operation |
| MLOAD/MSTORE | Memory operation |
| JUMP | Unconditional jump |
| JUMPI | Conditional jump |
| SLOAD | Storage operation |
| SSTORE | Storage operation |
| BALANCE | Get balance of an account |

### 6.2.4   Cost Estimation for the Escrow Deposit Simple-subgraph Pattern

We use the *Escrow Deposit* example of chapter 4 to show the gas cost estimation.  The simple subgraph for the Escrow Deposit, shown in Fig. 4.5, is repeated in Fig. 6.4 for readers' convenience.

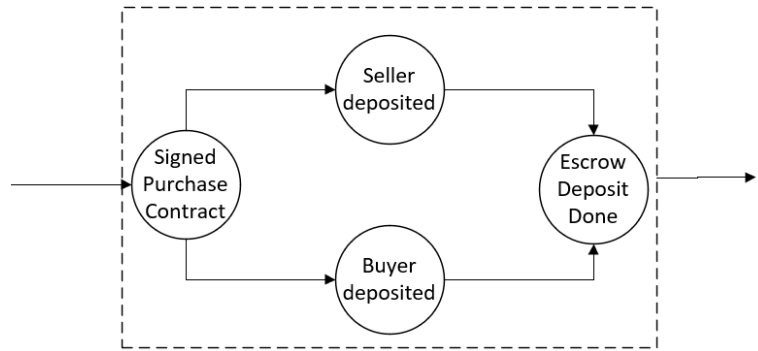

Fig.  6.4   The simple candidate subgraph for testing cost model

Please note that *seller deposited,* and *buyer deposited* states need to read on-chain data as they need to ascertain that the buyer and the seller did make their deposit, which is recorded on the blockchain. At the end of the pattern computation, upon the exit from the pattern, confirmation that the deposits were made has to be written to the blockchain.

The cost comparison will be conducted between

- the total cost of the simple-subgraph pattern executed only on the main chain

and

- the total cost of executing the same simple-subgraph pattern off-chain plus the overhead cost.

### 6.2.4.1  On-chain Execution Only

We can roughly calculate the gas consumption due to on-chain processing by assuming that each state has relatively the same size $M$ of state variables to read or update. Our specific Escrow Deposit example pattern has four states, where each of the states needs to load/store the data on-chain. Therefore,

The cost of computation of not moving to off-chain is:

$$On\text{-}chain\ only\ cost\ \ =\ \ 200(SLOAD)*4*M + 20000\ (SSTORE)*4*M = 80800\ M$$

### 6.2.4.2  Cost when pattern is executed off-chain

The off-chain pattern estimation is decomposed into the following components: (i) interface overhead; (ii) the on-chain computation before and after off-chain pattern; (iii) pattern's off-chain computation.

- Interface Overhead Cost Estimation

When off-chain processing completes, *attestCompute* function is invoked on-chain. However, we can conduct attestation through the variables in temporary memory or Stack. The only considerable gas consumption is on the updates of essential state properties by *offChainDoneEvent*. According to the interfaces we defined for the off-chain model in the previous chapter, we estimate gas consumption for each operation defined in the interface, as shown in Table 6.3. The size of the data accessed by the computation at each state is denoted as M.

Table 6.3    Interfaces Overhead Breakdowns

| | |
|---|---|
| begin { <br> wait for event *offChainEvent1*; <br>     if (state=='sen') OR *offChain* ) { <br>        *read blockchain data;* <br>        *assemble params and bc data;* <br>        *call off-chain processing w params;* <br>        assemble parameters from output; <br>     } <br>     if (state='sex') { <br>        *get bc data from cache;* <br>        *assemble bc data with parameters;* <br>        *boolean attested = attestResults (params)* <br>          *if NOT attested raise error Exception …* <br>     } <br> raise/fire *offChainDoneEvent1* event <br>     and pass it assembled parameters. <br> // if off-chain finished, params include bc data. <br> wait for *offChainEvent1*; <br> } | 10 (JUMP) <br> 200 * M (SLOAD: it is to bridge the *offchainevent*) <br> 3 *M (MLOAD/MSTORE) <br> (off-chain process invocation overhead) <br> 3 *M (MLOAD/MSTORE) <br> 10 (JUMP) <br> 3 *M (MLOAD/MSTORE) <br> 3 *M (MLOAD/MSTORE) <br> 3 *M (*attestResult* manipulates data in memory) <br> 10 (JUMP) <br> <br> 20,000 * M (SSTORE: *offchainDoneEvent* updates states variables in on-chain Storage) |

Consequently, the estimated cost of interfaces for *Escrow Deposit pattern* off-chain computation is:

$$gas\_consumption\_escrow\_deposit\_overhead$$
$$= 10 + 3\text{*}M + 3\text{*}M + 10 + 3\text{*}M + 3\text{*}M + 3\text{*}M + 10$$
$$= 30 + 15 * M$$

The interface overhead is directly proportional to the size of data required to be written to the blockchain upon completion of off-chain computation.

- Estimating the Cost of Pre- and Post-pattern Computation

We can also roughly calculate the gas consumption for on-chain computation before and after the off-chain pattern. Optimistically, the on-chain computation only happens on the entry and exit states of the pattern. The Signed Contract and Escrow Deposit Done states need to load/store the on-chain data. Therefore, we can roughly estimate the on-chain cost like

$$gas\_consumption\_escrow\_deposit\_on\_chain =200\text{*}2\text{*} M + 20000 \text{*}2\text{*}M = 40400\ M$$

- The Extra Off-chain Computation Cost

The example includes data communication with on-chain storge during the off-chain computation. Therefore, we should count the extra s communication cost for requiring on-chain data in off-chain computation into the total cost of off-chaining computation. Let us assume that off-chain computation needs to update the on-chain account balance of both buyer and seller. As a consequence, we need to count these as SSTORE operations twice. That is, we calculate the extra cost for on-chain data manipulation from off-chain computation as:

$$extra\_offchain\_computation\_cost = 200(SLOAD) * 2 * M + 20000 \ (SSTORE) * 2 * M = 40400 \ M$$

When the Escrow Deposit patter is executed off-chain, the total cost is the sum of on-chain computation cost, extra off-chain computation cost and the overhead of communication interfaces. As a result, the consumed gas will be:

$$Cost \ of \ computation \ off\text{-}chain = gas\_consumption\_escrow\_deposit\_overhead + gas\_consumption\_escrow\_deposit\_on\_chain + extra\_offchain\_computation\_cost = 30 + 15 \ M + 40400 \ M + 40400 \ M = 30 + 80815 \ M$$

## 6.3  DISCUSSION

The results show that this *Escrow Deposit* simple-subgraph pattern should not be moved off-chain because the total cost of moving to off-chain is more expensive than not moving to off-chain. However, the Escrow Deposit simple subgraph is just a particular use case in which off-chain pattern needs communication with on-chain storage. In most circumstances, the benefits brought by the off-chain model are obvious. For instance, let us assume that both states, Buyer Deposit and Seller Deposit, can be updated off-chain without communicating with on-chain data. The *extra_offchain_computation_cost* value is nearly zero because the off-chain event does not need to access on-chain data until the off-chain event is done. As a result, the saved cost can be:

$$gas\_saving = \ 80800 \ M - (30 + 15 \ M + 40400 \ M) = 40385 \ M - 30$$

The result shows that, generally, processing patterns off-chain can effectively save gas consumption. Again, Fig. 6.4 tells the fact that storing data on-chain storage is costly. We should store only data that is essential for computation and subsequent attestation when computation moves back to the mainchain and delegate storage of other data to other solutions, such as Swarm, Filecoin, IPFS (Aschenbach, 2018), and also move computation off-chain.   In short, we demonstrated that off-chain computation cost can be measured with the proposed estimation model. The purpose is to provide statistical evidence, beside the semantic information, to the blockchain application developers so they can make a more reliable decision about when to process a pattern off-chain.

It should be note that for simplicity, the cost estimation model does not include the frequency of execution of different methods that comprise a pattern – our model can be extended in a straightforward way to include that information.

# CHAPTER 7       CONCLUSIONS

## 7.1  LIMITATIONS

During the research, we recognized several limitations that we shall be addressing in our future work. Besides, offloading part of computations to off-chain raises the issues of increasing decentralization and additional security that will be required. We will continue research on how to move processing computation off-chain without compromising the beneficial properties of blockchains.

Our current cost model for off-chain computation does not expose the different frequencies with which the methods of a pattern, processed off-chain, may be invoked. The repetitive execution of the pattern's methods affects the benefits-overhead trade-off, and we are amending the cost model to account for them.

## 7.2  FUTURE WORKS

The scalability issue for complex blockchain applications, involving many actors, is an issue due to FSM modelling methodology due to the limited expressiveness of FSMs. Although HSMs reduce the problem, we will investigate decomposing subgraphs into individual actors' FSMs that cooperate together – a new approach in designing blockchain applications.

Lastly, the Layer 2 blockchain network solutions evolve rapidly nowadays. For instance, xDai sidechain network officially announced direct invocation between smart contracts defined on mainchain and sidechain network with integrated Oracle *Chainlink* interfaces in Aug 2020 (Geoknee, 2021). At the time of writing,  Robinson (2020) proposed an atomic *crosschain* transaction protocol for off-chain computation based on an up-to-date Layer 2 network solution. All these exciting changes give good opportunities for e-commerce type transactions to work efficiently on blockchains. Our potential research will be focused on modern techniques that have arisen.

## 7.3  CONCLUSIONS

We tackled the problem of executing parts of a smart contract computation off-chain. We adopt the separation of concerns to decompose the modelling process into solving the functional requirements first and only then we solve address which pieces of smart contracts can be processed off-chain and turn the model into a smart executable contract.

We employed FSM modelling to reflect the functional specifications, whereby the states of the FSM model are used to define an FSM graph. We use a graph algorithm to classify parts of the smart contract (parts of the FSM graph), referred to as simple subgraphs, that may be suitable to be executed off-chain. We proposed a model for assisting a developer in such decision making. While we gave just a few examples herein of our algorithm's implementation, we used it to identify several patterns, and the algorithm appears to be very useful for finding patterns. We analyzed the essential properties of the algorithm and found it intuitive enough for most developers to recognize simple subgraphs in an FSM graph only through visual inspection of the outcomes of the algorithm.

Once we locate the simple subgraphs patterns, these patterns are provided to the developer one at a time. With the pattern, the developer is also provided information on data input and produced by the pattern's methods as well as information on the cost overhead vs benefit. The developer reviews the provided information and determines if the pattern should be processed off-chain. We created a cost model to assisting the developer in evaluating the cost-benefit trade-off arising when a pattern is processed off-chain. We described how we adopt the transformation process (Mavridou & Laszka, 2018) to convert an FSM model into a smart contract. The method includes an API interface for the smart contract to communicate with off-chain computation. As overhead due to the interface, for communication between off and on-chain processing, is not negligible, we created a model of the interface that specifies the needed functionality that needs to be provided by the interface, so that the cost of such an interface can be estimated. The decision on whether to process a pattern off-chain, however, requires information that is of semantic nature in order to ensure that off-chain processing does not reduce trust in the blockchain application. We are investigating which semantic information is required and how to use it to make such a decision. Furthermore, we are adapting the transformation method of Mavridou & Laszka (2018(b)), which transforms an FSM model into a smart contract, so that the method would also provide an API interface for the smart contract to interact with off-chain computation.

In conclusion, while we delved into the problems that occur when patterns of a smart contract are processed off-chain, we recognized more work needs to be done, especially when determining if a pattern should be processed off-chain using the application semantic information that would mitigate the risks associated with lowering trust in the blockchain application when patterns are processed off-chain.

We assume that the power of smart contracts and blockchains will become a genuinely formidable and transformative power when models and tools are developed to reflect e-commerce type applications logic and when such models can be transformed into smart contracts without, or at least with limited, human intervention.